\documentclass[aps,prb,twocolumn]{revtex4}
\usepackage{amsmath,amsfonts,amssymb}
\usepackage{wrapfig}
\usepackage{graphicx}

\def \beq {\begin{equation}}
\def \eeq {\end{equation}}
\def \ba {\begin{eqnarray}}
\def \ea {\end{eqnarray}}
\newcommand{\upp}{\hspace{-2 pt}\uparrow}
\newcommand{\downn}{\hspace{-2 pt}\downarrow}
\newcommand{\SV}{\hat{\vec{S}}}
\newcommand{\AV}{\hat{\vec{A}}}
\newcommand{\IV}{\hat{\vec{I}}}
\newcommand{\EV}{\hat{\vec{E}}}
\newcommand{\Sz}{\hat{S}_z}
\newcommand{\Sy}{\hat{S}_y}
\newcommand{\Sx}{\hat{S}_x}
\newcommand{\Az}{\hat{A}_z}
\newcommand{\Iz}{\hat{I}_z}

\newcommand{\Sp}{\hat{S}_+}
\newcommand{\Ap}{\hat{A}_+}
\newcommand{\Ip}{\hat{I}_+}
\newcommand{\Sm}{\hat{S}_-}
\newcommand{\Am}{\hat{A}_-}
\newcommand{\IM}{\hat{I}_-}

\newcommand{\bra}[1]{\langle#1|}
\newcommand{\ket}[1]{|#1\rangle}

\newcommand{\ketbrad}[1]{|#1\rangle\!\langle #1|}

\newcommand{\mean}[1]{\langle#1\rangle}
\newcommand{\Mean}[1]{\left\langle#1\right\rangle}

\begin{document}
\title{Dephasing of quantum bits by a quasi-static mesoscopic environment$^{*}$}
\author{J.~M.~Taylor and M.~D.~Lukin}
\address{Department of Physics, Harvard University, Cambridge, MA 02138 USA}
\begin{abstract}
  We examine coherent processes in a two-state quantum system that is strongly
  coupled to a mesoscopic spin bath and weakly coupled to other
  environmental degrees of freedom.  Our analysis is specifically
  aimed at understanding the quantum dynamics of solid-state quantum
  bits such as electron spins in semiconductor structures and
  superconducting islands.  The role of mesoscopic degrees of freedom
  with long correlation times (local degrees of freedom such as
  nuclear spins and charge traps) in qubit-related dephasing is
  discussed in terms of a {\em quasi-static} bath.  A mathematical
  framework simultaneously describing coupling to the slow mesoscopic
  bath and a Markovian environment is developed and the dephasing and
  decoherence properties of the total system are investigated. The
  model is applied to several specific examples with direct relevance
  to current experiments. Comparisons to experiments suggests that
  such quasi-static degrees of freedom play an important role in
  current qubit implementations.  Several methods of mitigating the
  bath-induced error are considered. \\
$^*$\small{\emph{Dedicated to Anton Zeilinger, whose work has inspired
  exploration of quantum phenomenon in many avenues of physics and beyond.}}
\end{abstract}
\maketitle


\section{Introduction}

Solid-state quantum information research attempts the difficult task
of separating a few local degrees of freedom from a strongly coupled
environment.  This necessitates a clever choice of logical basis to
minimize the dominant couplings and environmental preparation through
cooling. Electron spin in quantum dots has been suggested as a quantum
bit~\cite{loss98,imamoglu99}, where orbital degrees of freedom do not carry
quantum information.  As a result, the individual quantum bits are
relatively well isolated from non-local degrees of freedom,
interacting strongly with only a local spin
environment~\cite{merkulov02,khaetskii02,desousa03b,taylor03,%
coish04,johnson05,koppens05,petta05} and weakly to phonons through
spin-orbit coupling~\cite{fujisawa01,golovach04,hanson03,%
elzerman04,johnson05}.  Josephson junctions and cooper-pair boxes
in superconducting systems can be engineered for decreased coupling to
local degrees of freedom~\cite{vion02}, and several different
approaches to superconductor-based quantum bit devices have shown
remarkable promise as quantum computation devices~\cite{pashkin03,%
chiorescu03,vion02,martinis02}.  However, superconductor designs
may strongly interact with local charge traps and magnetic
impurities~\cite{simmonds04,makhlin04,falci05}, leading to errors from local noise.

In this paper we examine the problem of coupling to the local
environment for a solid-state system.  By focusing on quantum
information-related tasks, the detrimental effect of local degrees of
freedom on specific operations can be assessed.  In our model,
so-called non-local degrees of freedom~\cite{stamp03} (such as phonons
and photons) are assumed to be weakly coupled and considered in the
normal Markovian limits.  However, local coupled systems corresponding
to spins, charge-traps, and other finite-level systems, can be
interpreted in terms of a finite set of nearby spins that may be
strongly coupled to the qubit.  In some cases, such as an electron
spin in a quantum dot, this heuristic picture corresponds exactly to
the actual hyperfine coupling between electron spin and lattice
nuclear spins.  In those systems, the local environment consists of
$10^{4}$--$10^6$ lattice nuclear spins, and the coupling can be
strong~\cite{merkulov02,khaetskii02,taylor03,taylor03b}.  For
superconductor-based designs, the quantum bit couples both to actual
spins and to charge degrees of freedom such as $1/f$-type
fluctuators~\cite{weissman88}, which can be modeled as two-level
systems.  Thus, analysis of the most general coupling of a qubit to a
mesoscopic collection of spins yields an understanding both of
limitations in current experiments and of methods for improving the
operation and design of solid-state quantum computation devices.  We
remark that several works have analyzed this type of situation in
detail for specific implementations, in superconducting
qubits~\cite{falci05,makhlin04} and in quantum dot
qubits~\cite{desousa03b,coish04,coish05,hu05,klauser05}.  We also note
that recent work focusing on quantum dot-based systems parallels
several significant elements of the current paper~\cite{klauser05};
the results were arrived at independently.  The present work focuses
on the generality of the model and its impact in the context of
quantum information processing.

We begin by considering a spin-1/2 system (the ``qubit'') coupled to a
finite number of spin-1/2 degrees of freedom (the ``mesoscopic bath'')
to both weak and strong driving fields.  Inclusion of weaker,
non-local couplings (the ``Markovian environement'') through a
Born-Markov approximation reveals a natural hierarchy.  A separation
of time scales indicates that the local environment is quasi-static
with respect to experimental time scales with long-lived correlation
functions.  However, over many experimental runs, the local
environment fluctuates sufficiently to defy efficient characterization
and correction by means of direct measurement.  The generic nature of
the coupling allows bath characterization using only a few parameters.
Then, detailed analysis of specific operations (phase evolution, Rabi
oscillations, spin-echo) and comparison to experimental results is
possible.  While such a spin-bath model has been considered
previously~\cite{zurek81,prokofev00,rose01,merkulov02,makhlin04,falci05},
the significance of the mesoscopic and quasi-static limits has not
been emphasized.

We find that a quasi-static, mesoscopic environments results in fast
dephasing, induces power-law decay and non-trivial
phase shifts for driven (Rabi) oscillations of the qubit, and may be
corrected for using echo techniques.  The generality of the model
suggests it may be appropriate for many systems with strong coupling
to a stable local environment.
In addition, comparison to experiments suggest this model can help
explain the reduced contrast and long coherence times of current
solid-state qubit systems, indicating that the local environments of
current solid-state qubits both have long correlation times and play a
crucial role in the dynamics of the qubit.  Several methods of
mitigating the effects of the bath are considered, all of which take
advantage of the long correlation time to reduce errors.  Some of
these techniques are extensions of established passive error
correction~\cite{zanardi97} and quantum bang-bang
ideas~\cite{viola98}, while another, the preparation of coherences in
the environment to reduce dephasing (environmental squeezing) is
introduced in this context for the first time.

\section{The physical system}
Consider a relatively well-isolated two-level system (a qubit) in a
solid-state environment.  This could be a true spin-1/2 system such an
electron spin or an anharmonic system such as a hybrid flux-charge
superconductor qubits~\cite{foot1}.
The qubit is coupled to a collection of spin-1/2 systems comprising
the mesoscopic bath.  In addition to these potentially strongly
coupled degrees of freedom, other, weaker couplings to bosonic fields
(phonons, photons, and large collections of spins that can be mapped
to the spin-boson model~\cite{feynman63}) are included as a Markovian
environment, and act as a thermal reservoir.  Qualitatively, this
leads to a heirarchy of couplings.  The qubit plus mesoscopic bath are
treated quantum mechanically, while the additional coupling to the
larger environment (the thermal reservoir) is included via a
Born-Markov approximation.  Thus, the reservoir plays the role of
thermalizing the mesoscopic bath over long time scales and providing
additional decay and dephasing of the system.  By assuming slow
internal dynamics and thermalization for the mesoscopic bath, a
quasi-static regime is investigated, in contrast with the usual
Markovian approximations for baths.

Starting from first principles, the hamiltonian is introduced.  A
transformation to the rotating frame allows for adiabatic elimination
which simplifies the interaction, after which the quasi-static
assumption is considered.  An explicit bath model is chosen, and the
role of the thermal reservoir included through a Linblad-form
Louivillian.

\subsection{Hamiltonian}

The qubit, 2-level system, is described by the spin-1/2 operators
$\Sx, \Sy, $ and $\Sz$.  An external biasing field produces a static
energy difference between the two levels of angular frequency
$\omega$.  This serves as a convenient definition of $z$-axis.  A
driving field, with Rabi frequency $\Omega$ and oscillating with angular frequency
$\nu$, is applied along a transverse axis in the $x-y$ plane.  In most
systems, $\omega$ is fixed, and modulation of $\Omega$ turns on and off
rotations around the transverse axis.  The relative phase of the
driving field to the biasing field controls the axis about which
rotations occur.  We choose the phase such that rotations occur around
the $x$-axis.  The hamiltonian is
\beq
\hat H_{\rm sys} = \omega \Sz + 2 \Omega \cos ( \nu t) \Sx ,
\eeq
in units where $\hbar = 1$.  This hamiltonian is commonly encountered
in quantum optics, NMR, ESR, and solid-state quantum information
devices.  For solid-state qubit systems, $\omega$ and $\nu$ are
typically $\sim 1-100$ ns$^{-1}$ and $\Omega \sim 0.01-1$ ns$^{-1}$.

Adding the most general environmental coupling possible, the total hamiltonian is
\beq
\hat H_{tot} = \hat H_{\rm sys} +  \tilde \lambda \SV \cdot \EV + \hat H_E
\eeq
where $\tilde \lambda$ is a coupling constant to the environment and $\EV$ is a vector of environmental operators.  The tilde terms indicate that this coupling includes both the mesoscopic spin bath and the Markovian environment. All of the internal environmental dynamics, including thermalization, is encompassed in $\hat H_E$.

\begin{figure}
\begin{center}
\includegraphics[width=3.0in]{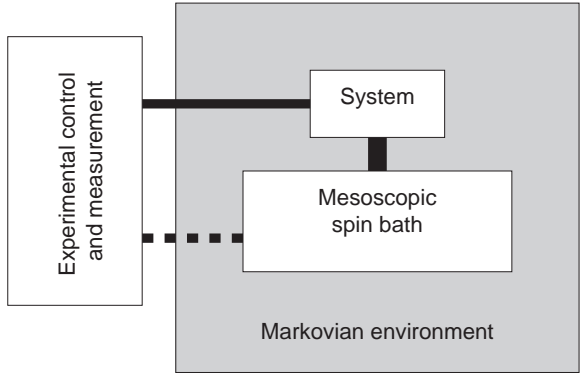}
\caption{
Heirarcichal system--bath--environment coupling.  External control of the system (qubit) and measurement are independent of the environment
The mesoscopic spin bath is strongly coupled to the system, while the environment provides weak dephasing of the system and spin bath degrees of freedom.  Internal dynamics of the spin bath can be ergodic, but some experimental control of the bath may be achieved in limited situations (e.g., through NMR) as indicated.
\label{f:heirarch}}
\end{center}
\end{figure}

Separating the environment E into the mesoscopic spin bath and
Markovian environment, the interaction and environment terms in the
hamiltonian can be rewritten:
\beq
 \tilde \lambda \SV \cdot \EV + \hat H_E =  \lambda \SV \cdot \AV +  \lambda_C \SV \cdot \hat{ \vec{C}} 
  + \hat H_A + \hat H_C + \hat H_{AC} .
\eeq
The $A$ terms denote the quasi-static bath degrees of freedom, while
the $C$ terms correspond to a larger environment that can be treated
in a Born-Markov approximation. Qualitatively, the complete picture of
bath and environment drawn in this paper is hierarchical in nature
(Fig.~\ref{f:heirarch}), with the system (qubit) coupled to a
mesoscopic spin bath, both of which are coupled to larger
environments.

This paper focuses on the regime of strong system--bath coupling and
weak system--environment and environment--bath coupling.  In this
limit, we can treat the Markovian environment's effects entirely
through a Lindblad form Louivillian acting on the combined
system--bath.  We write the effective hamiltonian as
\beq
\hat H_{\rm eff} = \hat H_{\rm sys} + \lambda \SV \cdot \AV + \hat H_A\ ,
\eeq
while superoperator describing evolution of the system--bath density
matrix $\hat \rho$ is given by the differential equation
\beq
\dot \hat \rho = i [ \hat \rho, \hat H_{\rm eff} ] + {\mathcal L}_{\rm env} \hat \rho\ . \label{e:rhodot}
\eeq
The Louivillian, ${\mathcal L}_{\rm env}$, is described below.

Regarding other limits, the weak system-bath limit can be treated
pertubatively when the bath-environment coupling and/or
system-environment coupling is stronger.  When the system-environment
coupling can be experimentally controlled, such as during qubit
manipulation,  the system degree of freedom could be used to
transfer entropy from the bath to the environment in a controlled
manner, yielding non-thermal initial states of the mesoscopic spin
bath.  The usefulness of this final point will be discussed later.

In many cases, the bias field $\omega$ is large ($| \omega | \gg |
\Omega |, |\lambda|$).  This is similar to the quantum optical case;
accordingly, the standard recipe of transformation into a rotating
frame and a rotating wave approximation is also valid here, and
off-axis (flip-flop) terms are removed, resulting in a
Jaynes-Cummings-like hamiltonian.  Roughly speaking, energy cannot be
conserved when the system spin flips through changes of the bath
state.  However, the additional coupling to the bath leads to terms
not found in the Jaynes-Cummings hamiltonian.

More explicitly, applying the unitary transformation $U = \exp(-i \omega \Sz t)$, the hamiltonian is
\ba
\hat{\tilde{H}}_{\rm eff} & = & \Omega / 2 [ e^{i \nu t} + e^{-i \nu t}] [ e^{i \omega t} \Sp + e^{-i \omega t} \Sm ] +  \nonumber  \\
& & \lambda \Sz \Az + \lambda / 4 [ e^{i \omega t} \Sp \Am + e^{-i \omega t} \Sm \Ap ] + \nonumber  \\
& & \hat H_A .
\ea
When the detuning of the driving field is small, $\delta = (\nu -
\omega) \ll \omega$, the highly oscillatory terms in this rotating
hamiltonian can be neglected.  To do so, the propagator $\hat U(t) =
\hat T \exp(-i \int_0^t \hat{\tilde{H}}(t') dt')$ is formally expanded
over one period of the applied unitary transformation, $\tau = 4 \pi /
\omega$, and a Magnus expansion~\cite{magnus54} gives
\ba
\hat U(\tau) &=& \exp[- i \tau ( \sum_{k=0}^{\infty} \hat H_k) ]  \\
\hat H_0 & = & 1/\tau \int_0^{\tau} \hat{\tilde H}(t)~dt \nonumber \\
  & = & \Omega/2 [ \Sp + \Sm ] + (\delta + \lambda \Az) \Sz  + \hat H_B \label{e:hnot} \\
\hat H_1 & = & -i/2 \tau \int_0^{\tau} \int_0^{t_1} [\hat{\tilde H}(t_2), \hat{\tilde H}(t_1)]~dt_2~dt_1 . \label{e:hone}
\ea
The first term $\hat H_0$ will be used hereafter.  Neglecting $\hat H_1$ and higher terms of the expansion if formally equivalent to the rotating wave approximation used to derive the Jaynes-Cummings hamiltonian.

\subsection{Quasi-static bath assumptions}

In standard analyses, usually the bath dynamics are fast relative to
the system dynamics, and a Born-Markov approximation is appropriate.
However,  when the internal bath dynamics are much slower than the
system dynamics, the internal bath correlation functions are long
lived, and we can consider a {\em quasi-static} bath.  In this
picture, one collective bath operator (here $\hat{A}_z$) has, in the
Heisenberg picture, a long-lived correlator.  In particular,
\beq
1- \frac{\mean{\Az(t+\tau) \Az(t)}}{\mean{\Az^2}} \ll 1
\eeq
for $\tau \gtrsim 1/\Omega$.  The quasi-static limit is approached
when this is satisfied for $\tau > t_{\rm expt}$ (the time for a
single experimental run) but not for $\tau \simeq t_{\rm tot}$ (the
time to generate enough runs to characterize $\Az$).

We consider this in detail. Internal dynamics of the spin bath and
coupling to the Markovian environment will result in a decay of
correlations.  A general description of the correlator is
\beq
\mean{ \Az(t) \Az(t')} = \int d\omega\ S(\omega) e^{i\omega(t-t')} \label{e:spectralFcn}
\eeq
where the spectral function $S(\omega)$ has some high frequency
cutoff, $\Gamma$, set by the internal dynamics of the bath~\cite{cottet01}.  When
$\Gamma \gg \Omega$, the Markovian limit may be reached, and the
quasi-static theory developed here is no longer necessary.

When $\Gamma \rightarrow 0$, the thermalization time goes to infinity.
Then, while the state of the bath is initially unknown, a series of
measurements can be used to estimate the detuning of the system, and
thus the value of $\lambda \Az$ (see, {\em e.g.},
Refs.~\onlinecite{giedke05, klauser05}).  If the manipulation process has
a measurement time for a single projection of $\tau_m \gtrsim
1/\Omega$, after $n$ such measurements, the error of $\mean{\Az}$ will
be
\beq 
\Delta({\mean{\Az}}) \simeq \frac{\lambda }{ \Omega \sqrt{n}} .
\eeq
This procedure is exactly that of a classical phase estimation
algorithm.  It takes a time $\propto n$ to generate a $\sqrt{n}$
improvement in the knowledge of the current value of $\Az$.

In between lies the quasi-static limit, where a single experiment
({\em i.e.}, preparation of a given qubit state, evolution under $H_0$ for a
specified time, followed by projective measurement) can be performed
well within the correlation time, but not enough experiments can be
performed to well estimate $\Az$.  Qualitatively speaking, this is
equivalent to taking a thermal average for the ensemble of
measurements, and mathematically the same as the result for the
simultaneous measurement of a large number of exactly similar systems.
The uncertain knowledge of $\Az$ leads to dephasing when attempting
specific quantum information operations.

As a typical case, consider an internal bath hamiltonian $H_A = \hbar
\sum_k \omega_{B,k} \hat{B}^k_z$ and a coupling with $\AV = \sum_k
\alpha_k \hat{\vec{B}}^k$.  This could describe nuclear spins for
quantum-dot systems, or charge traps.  As $[\hat H_A,\hat{A}_+] =
\sum_k \alpha_k \omega_{B,k} \hat B_+^k$, the rotating-wave
approximation is valid if $\omega_{B,k} \ll \omega$ for all $k$.
Then, the application of a large bias field $\omega$, e.g., due to a
static magnetic field, makes the interaction essentially static and
classical, whereby the system experiences a fluctuating (classical)
field $\AV$; fluctuations are due to corrections to the rotating wave
approximation and coupling to the Markovian environment.
Non-classical correlations of the bath, produced by off-resonant
interactions~\cite{taylor05prb,klauser05} or preparation techniques
(e.g.~\cite{taylor03b}) return the problem to the fully quantum
domain.  

For example, for electron spin in quantum dots, where the bath is
lattice nuclear spins, dipolar diffusion processes lead to flip-flop
interactions wherein a spin inside the dot is exchanged with that of a
spin outside the dot, producing ergodic dynamics within the bath on ms
timescales~\cite{desousa03,yao05}.  In contrast, spin-lattice
relaxation (coupling to the larger environment) can be minutes to
hours in time.  Over time, the dipole diffusion destroys the
correlations of the $\Az$ parameter.  Detailed calculations, using the
method of moments, suggest that diffusion in dots is within an order
of magnitude of the bulk diffusion rate~\cite{deng03}, and this
diffusion decorrelates $\Az$ no faster than the time scale set by the
NMR linewidth (10 ms$^{-1}$ for GaAs~\cite{paget83}).  Thus the
primary effect of spin-lattice relaxation (coupling to a Markovian
environment) is to produce an initial thermal state of nuclear spins,
while dipole-dipole interactions lead to decorrelation of $\Az$
(internal bath dynamics).

For concreteness, we choose for our mesoscopic spin bath of $N$
spin-1/2 systems to have the form 
\beq
\Az = \sum_k \alpha_k \Iz^k \label{e:azdef}
\eeq
where $\Iz^k$ is the spin component corresponding to the $k$th bath
spin, and an internal Hamiltonian giving the quasi-static limit.  As
many bath variables scale as $\sqrt{N \bar{\alpha^2}}$, we normalize
the $\alpha_k$ such that $\bar{\alpha^2} = 1/N$, thus keeping
$\lambda$ as the sole bath strength parameter.  The bars above
quantities denote statistical averages over static variables.  As we
have assumed all bath spins are spin-1/2, we can choose the $z$ axis
without loss of generality.

\subsection{Markovian environment effects}

The principle effects of the Markovian environment are two-fold: it
leads to relaxation of the qubit system, and it leads to long-time
thermalization of the mesoscopic spin bath.  We have established that
the internal dynamics of the quasi-static bath, combined with coupling
to the Markovian environment, will lead to decorrelation of $\Az$ and
dynamic thermalization.  For a weakly coupled Markovian environment,
with equilibriation times for the bath that are much longer than
experimental times, the main effect of the Markovian environment is to
produce an initial mesoscopic spin bath state of the form $\rho_A
\propto e^{-H_A /k_{\rm B} T}$.

In general, the additional Markovian environment degrees of freedom
will also couple to the system, leading to relaxation and decoherence
that we can model with a Lindblad form of the Louivillian of the combined
system-spin bath density matrix.  Formally, tracing over the
environment yields a superoperator (see Eqn.~\ref{e:rhodot})
\ba
{\mathcal L}_{\rm env} \rho & = & 2 \gamma_s [2 \Sm \rho \Sp -  \rho \Sp \Sm - \Sp \Sm \rho ] \nonumber \\
& & + \sum_k 2 \Gamma_k \Big[ 2 \IM \rho \Ip -  \rho \Ip \IM - \Ip \IM \rho \nonumber \\
& & + 2 \Ip \rho \IM - \rho \IM \Ip - \IM \Ip \rho \Big]
\ea
corresponding to the Markovian environment.  We presume that the
spin-bath is in the high temperature limit, while the qubit system
only has spontaneous decay.  This is consistent with the rotating wave
approximation.  The resulting equations, with pure radiative decay for
the system, are cumbersome to work with, and we will sometimes use the
input-output operator formalism, which is formally equivalent by the
quantum fluctuation-dissipation theorem.

\section{Free evolution}

In this section, free evolution for a variety of
experimentally-relevant situations is considered.  Free evolution
corresponds to the case of $\Omega=0$, {\em i.e.}, the absence of driving
field.  In quantum information, free evolution is equivalent to phase
rotations (Z rotations), and the behavior of free evolution is an
indicator of how good a quantum memory the chosen qubit provides.  The
decay of coherence during free evolution due to the mesoscopic spin
bath may be severe; however, we consider spin-echo methods as a means
of mitigating the bath's effects.

Equation~\ref{e:hnot} is the basic Hamiltonian used hereafter; $\Az$
is given by Eqn.~\ref{e:azdef}.  For free evolution, $\Omega=0$, and
the Hamiltonian is
\beq
\hat H_{FID} = (\lambda \Az + \delta) \Sz \ .
\eeq
For time-dependent $\Az$ such that $[\Az(t),\Az(t')]=0$ (appropriate for the specific baths consider in section II), the exact propagator from $t=0$ to time $t$ is
\beq
U_{FID}(t) = \exp[-i (\delta t + \lambda \int_0^t \Az(t') dt')\Sz]\ .
\eeq
In the quasi static limit, we may usually replace $\int_0^t \Az(t') dt'$ with $\Az t$, simplifying evaluation of the propagator.

\subsection{Free induction decay}
To illustrate free evolution, we start by studying free induction
decay (FID), considered for this case by~\cite{zurek81} and for a
dynamical spin bath by~\cite{teklemariam03}.  FID corresponds to the
decay experienced by a spin coherence (e.g., a density matrix element
$\ket{\upp} \bra{\downn}$) during evolution in a Ramsey-type
experiment.  The system is prepared in the state $\ket{+} = 1/\sqrt{2}
( \ket{\upp} + \ket{\downn} )$, with $\mean{ \Sp(0) } = 1/2$, for
example, by rotations induced by control of $\Omega$ (considered
below).  The driving field is turned off ($\Omega = 0$) and the system
evolves according to the $\Sz$ coupling only.  Then,
\beq
\Phi_{FID} = \frac{\mean{ \Sp(t) }}{\mean{\Sp(0)}} = \Mean{ \exp[-i (\lambda \Az + \delta) t ] }_{bath}
\eeq
For an uncorrelated bath ($\mean{ \Iz^k \Iz^j } = \mean{\Iz^k}
\mean{\Iz^j}$ for $k \neq j$) we can write this in terms of individual
bath couplings\cite{foot2}
\beq
\Phi_{FID} = e^{-i \delta t} \prod_{k} \big[ \cos (\lambda \alpha_k t / 2) - i \mean{ 2 \Iz^k}  \sin (\lambda \alpha_k t / 2) \big] .
\eeq
The minimum of the projection of the actually final state of the
desired final state, $F_{FID} = | \Phi_{FID} |^2$ is bounded from
below by $({\rm Min}_k [ |\mean{ 2 \Iz^k} | ])^N$.  Thus, initial bath
polarization for finite $N$ limits the maximum decay.

\begin{figure}
\begin{center}
\includegraphics[width=3.0in]{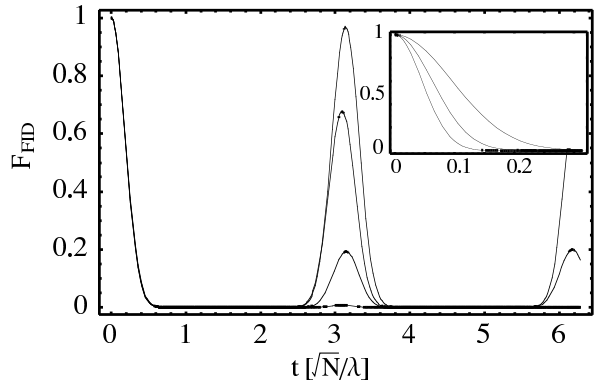}
\caption{ 
FID evolution ($F_{FID}$) in time scales $\sqrt{N}/ \lambda$ for increasing inhomogeneity, $\sigma_{\alpha} = \{0.1,0.3,0.6,1\}/\sqrt{N}, N=30,P=0$.  The inset shows
increasing decay in the short time limit due to increasing inhomogeneity, $\sigma_{\alpha} = \{5,10,20\}/\sqrt{N}, N=30, P=0$.
\label{f:FID}
}
\end{center}
\end{figure}

The initial decay of this coherence is quadratic, giving
\beq
F_{FID} \simeq 1 - \frac{(\lambda t)^2}{2} \sum_k 2 \alpha_k^2 ( 1- \mean{2 \Iz^k}^2) ,
\eeq
as shown in the inset of Figure~\ref{f:FID}.
For a large number of spins, the intermediate time decay converges to
a Gaussian, $\exp[ - (\gamma_{FID} t)^2/2]$, with a rate 
\beq
\gamma_{FID} = \lambda \sqrt{\sum_k \alpha_k^2 ( 1- \mean{2 \Iz^k}^2)} \label{e:gammaFID}.
\eeq

When the inhomogeneity in the $\alpha_k$ coefficients is low, the
system experiences mesoscopic revival on a time scale given by the
single spin coupling strength, $\lambda \bar \alpha \simeq
\lambda/\sqrt{N}$.  This type of revival is shown in
Figure~\ref{f:FID}, where a Gaussian distribution of $\alpha_k$ of
increasing widths are used, and a bath polarization $P = \mean{2 \Iz}
= 0$ is assumed.  These mesoscopic effects for finite spin systems
have been investigated in great detail elsewhere~\cite{zurek81}, and
we refer readers to those works for a detailed description.  In most
physical settings, inhomogeneity is large and revival will not be
observed.  However, for systems where $N$ is small, fluctuations can
become substantial even with a large inhomogeneity in the
$\alpha_k$'s, and such fluctuations have been observed
experimentally~\cite{teklemariam03}.

To put the results in an experimental context, we consider a Ramsey
fringe experiment.  A system is prepared in the $\ket{\downn}$ state.
A $\pi/2$ pulse is applied along $x$ (denoted $R^x_{\pi/2}$), and then
at a later time $t$, an opposite, $-\pi/2$ pulse is applied.
Afterwards, a projective measurement on $z$ measures the rotation of
the $\Sy$ eigenstate due to environmental effects and detuning.  In
terms of actual experimental implementations, preparation of a
spin-down eigenstate and measurement along the $\Sz$ axis is enough,
if we use two $\pi/2$-pulses with a delay $t$.  Formally, in the limit
of perfect pulses the propagator is
\beq
U_{FID}(t) = R^x_{-\pi/2} \exp[-i \lambda \Az \Sz t ] R^x_{\pi/2} . \label{e:FIDprop}
\eeq
In the limit of perfect pulses, the probability of measuring the final
state in the $\ket{\downn}$ state is given by $F_{FID}$ as a function
of the delay time, $t$.  If the system is coupled to a spin bath, it
will show Gaussian-type decay in the initial time period.  This decay
is limited in extent for baths of finite polarization (non-infinite
temperature).  For a small bath size, mesoscopic fluctations and
revival after the initial decay may be evident.

For a series of sequential measurements, the finite correlation time
of the bath will be evident.  For example, for fixed interaction time
$\tau$, a series of measurements  $\{ M_k \}$ giving 1 for
$\ket{\downn}$ and 0 for $\ket{\upp}$ at times $t_k$ will manifest
correlations due to the correlations of $\Az$ at the different times,
$t_k$.  Assuming the quasi-static limit, such that $\hat M_k =
\cos^2(\lambda \Az(t_k) \tau/2)$, we find
\beq
\mean{M_j M_k} - \mean{M}^2 = \frac{1}{8} \exp[-(\lambda \tau)^2 \int d\omega S(\omega) (e^{i \omega (t_j - t_k)} - 1)]
\eeq
where the cutoff $\Gamma$ of $S(\omega)$ is such that $\Gamma \tau \ll
1$.  Thus, the correlation function for a series of measurements with
$\tau \Gamma \ll 1$ in principle allow reconstruction of the spectral
function describing $\Az$.  We note that $\gamma_{FID} = \lambda
\sqrt{\int S(\omega) d\omega}$.

\subsection{Spin-echo decay}
As the bath correlation time is long, 
spin-echo techniques can exactly cancel this type of dephasing. For spin-echo, a $\pi$-pulse is applied in between the two $\pi/2$-pulses; the time delays before and after the $\pi$-pulse are $t_1$ and $t_2$, respectively.  The total evolution is thus
\ba
U_{SE}(t_2;t_1) &=& R^x_{\pi/2} \exp[-i \lambda \Az \Sz t_2 ] R^x_{\pi} \nonumber \\
 & & ~\exp[-i \lambda \Az \Sz t_1 ] R^x_{\pi/2} \\
& = & R^x_{-\pi/2}\left( R^x_{-\pi} \exp[-i \lambda \Az \Sz t_2 ] R^x_{\pi}\right) \nonumber \\
& & ~\exp[-i \lambda \Az \Sz t_1 ] R^x_{\pi/2} \\
& = & R^x_{-\pi/2} \exp[-i \lambda \Az \Sz (t_1-t_2)] R^x_{\pi/2} ,
\ea
which is the original FID propagator (Eqn.~\ref{e:FIDprop}) with $t \rightarrow (t_1-t_2)/2$.  The previous results hold but now as a function of the time difference.  In particular, no decay occurs when $t_1 = t_2$!  The probability of measuring $\ket{\downn}$ is
\ba
F_{SE}(t_1,t_2) &= & \mean{\cos^2 [\lambda \Az (t_1-t_2)/2]} \nonumber \\
 &= &\frac{1}{2}\{ 1 + \exp[-\gamma_{FID}^2 (t_1-t_2)^2/2] \}.
\ea

Spin-echo fidelity ($F_{SE}(t,t)$) is limited both by the imperfections in the rotations, $R^x_{\pi/2}$ and $R^x_{\pi}$,  Markovian processes that directly lead to decay of $\Sz$, and decay of the correlation function of $\Az$, {\em i.e.}, in the interaction picture, $\Az(t) \neq \Az(t')$ for $t$ much later than $t'$.  The first type of error, examined in detail below, does not depend on the total time $t_1 + t_2$, while the latter errors do. Neglecting imperfections in rotations, 
the effect of a correlated bath on spin-echo is a textbook problem, and we merely cite the result here:
\ba
F_{SE}(t,t) &=& \frac{1}{2} + \frac{e^{-2 \gamma t}}{2} \\
& & \ \  \times \exp[-\lambda^2 \int S(\omega) \sin^4(\omega t/2)/(\omega/2)^2 d\omega] \nonumber
\ea
where we assume $\Az$ is a Gaussian variable described by the spectral function $S(\omega)$, as per Equation~\ref{e:spectralFcn}.  The results of a spin echo experiment allow an alternative method of extracting the correlation function associated with the mesoscopic spin bath. For a spectral function describing the bath with cutoff $\Gamma \ll 1/t$, the echo will decay according to $\exp[-2 \gamma t - \lambda^2 \Gamma^2 t^4/4]$.

For free evolution ($\Omega=0$), used for phase gates and quantum memory, we find that a mesoscopic spin bath leads to dephasing in the quasi-static limit, with a time scale given by $1/\lambda$.  Sequential measurements in a Ramsey-type experiment allow for reconstruction of the of the spectral function associated with the bath. Furthermore, spin-echo can be used to reduce the dephasing dramatically.  

\section{Driven-evolution}

While the free evolution of the system is illustrative of quantum memory, a critical operation for quantum information processing is rotations about an axis perpendicular to the applied bias.  This is achieved by a pulse in the transverse coupling field strength, $\Omega$, leading to driven oscillations between $\ket{\upp}$ and $\ket{\downn}$.  These Rabi oscillations form the basis for the $x$-axis rotations used, for example, in Ramsey-fringe experiments, and more generally for producing a universal set of single qubit gates.

In practice, the perfect rotations (Rabi pulses) needed to produce perfect rotations are unavailable, due to the finite power available for such pulses.  We now consider in detail rotations obtained by driving the system (non-zero $\Omega$ in Eqn.~\ref{e:hnot}) in the presence of the mesoscopic spin bath and Markovian environment.  In the Heisenberg picture, the single-spin (system) operator equations of motion are given by
\beq
\frac{d\SV}{dt} = i [ H_0 , \SV ] .
\eeq
For time independent $\Az$ (quasi-static limit) we can solve these equations exactly, noting that $d/dt~\SV= \hat \omega \vec{n} \times \SV$, where
\begin{eqnarray}
\hat{\omega} & = & \sqrt{(\lambda \Az + \delta)^2 + \Omega^2} ;  \label{e:omega} \\
\hat{\vec{n}} & = &  \hat{\omega}^{-1} \left( 
\begin{array}{c}
\Omega \\
0 \\
\lambda \Az + \delta
\end{array} \right) \label{e:vectors} .
\end{eqnarray}
This evolution corresponds to the optical Bloch equations, giving driven rotations of the qubit Bloch vector on the Bloch sphere about an axis $\vec{n}$ at frequency $\hat \omega$.  The dependence these quantities upon the state of the bath leads to shot-to-shot fluctuations, which broadens ensemble or sequential measurements.

Alternatively, we may write the propagator for $\hat H_0$ as
\beq
\hat U(t) = \cos(\hat \omega t/2) \hat{1} - 2 i  \sin(\hat \omega t/2) \hat{\vec{n}} \cdot \SV
\eeq
where we have taken the quasi-static limit, such that $\frac{d}{dt} \Az \simeq 0$.

Returning to Equations~\ref{e:omega}--\ref{e:vectors}, we calculate the rotation of an initial $z$ eigenstate with $\mean{\Sz (t=0)} = 1/2$, {\em i.e.}, determine the Rabi oscillations expected for the initial state $\ket{\upp}$.  The field $\Omega$ leads to population and coherence oscillations.  Considering the $z$ and $y$ components together, we define
\ba
\hat f &=& \frac{(\lambda \Az + \delta)^2}{(\lambda \Az + \delta)^2 + \Omega^2} \\
\hat \zeta(t) &=& \frac{\Omega^2 \exp(-i \sqrt{(\lambda \Az + \delta)^2 + \Omega^2} t)}{
(\lambda \Az + \delta)^2 + \Omega^2} .
\ea
Then, $\mean{ 2 \Sz(t) } = {\rm Re}[\mean{ \hat f + \hat \zeta(t) }]$ and $\mean{ 2 \Sy(t) } = {\rm Im}[\mean{\hat \zeta(t) }]$.  The value of $\mean{\hat f}$ is a measure of the maximum contrast of population oscillations, and also gives the steady-state population difference and lineshape; this is approximately maximal for $\delta = - \lambda \mean{ \Az}$.  $\mean{\hat \zeta(t)}$ gives the oscillatory part of coherence and population.

\subsection{Steady-state behavior}

For long times, the steady state behavior will be given by $\mean{\hat f}$, which measures the population difference between $\upp$ and $\downn$.  This gives the measureable response of the system to CW excitations of the system.  A simple estimate of $\mean{\hat f}$ is provided by writing $\Az$ as $\mean{\Az} + \delta \Az$, and assuming $\mean{(a + \Az)^{-2}} = \mean{(a + \mean{\Az})^2 + \mean{\Delta \Az^2})^{-1}}$.  Then, 
\beq
\mean{\hat f}= \frac{1}{1+\Omega^2/[(\delta +\lambda \mean{ \Az})^2 + \lambda^2 \mean{\Delta \Az^2}]}.
\eeq
This has a maxima at the mean bath detuning, $\lambda \mean{\Az}$, and as such this point would be the observed zero in detuning.  Differences from this observed zero, denoted $\tilde \delta$, gives an absorption spectrum of
\beq
\mean{\hat f} =1 -  \frac{\Omega^2}{\tilde \delta^2 + \lambda^2 \mean{\Delta \Az^2} +  \Omega^2} \label{e:roughline}
\eeq
which behaves as a Lorenztian with a linewidth given by the combined power broadening and bath broadening for $\Omega \ll \lambda \mean{\Delta \Az^2}^{1/2}$.  

The approximations used for this simple estimate break down for large bath strength.  To find this behavior in the strong bath limit, where higher order moments are important, we need not assume that the bath density matrix is uncorrelated, as before, but rather that the 
bath density matrix is diagonal in an eigenbasis of the bath operator,
$\Az$.~\cite{foot3}
This would be appropriate for nuclei interacting with an electron spin in a quantum dot at finite magnetic field, or for charge traps capacitively coupled to the system.  

\begin{figure}
\begin{center}
\includegraphics[width=3in]{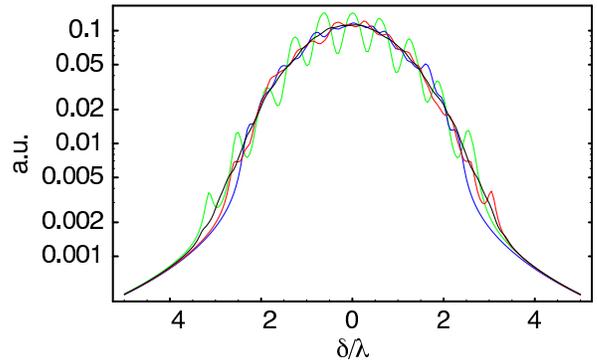}
\caption{Lineshapes for a bath with $N=10$ and increasing inhomogeneity [$\sigma_\alpha$ = 0.1(green),0.5(red),1(blue)].  
The additional line is $N=20,\sigma_{\alpha} = 0.5$ (black).}
\label{f:lineshape}
\end{center}
\end{figure}

We can evaluate $\mean{\hat f}$ by tracing over the bath density matrix, replacing $\rho_{bath} = \sum_{\lambda A_z} \rho_{\lambda A_z} \ketbrad{\lambda A_z}$ with $\rho(\Lambda)$.  The integrals involved are solved in Appendix~\ref{s:math}. We note that the result is 
\beq
\mean{\hat f} \simeq 1 - \rho(-\delta) \sqrt{2 \pi \Omega/u} \label{e:lineshapeReal}
\eeq
for the lineshape.  This has assumed that $u^{-1}$ is a frequency scale over which the bath density of states is relatively flat, and as such can be Taylor expanded to second order.  This time scale is
\beq
u = \frac{5}{4 \Omega} -\frac{\Omega \rho''(-\delta)}{\rho(-\delta)} . \label{e:u}
\eeq
In other words, driven oscillations can measure the bath density of states as a function of detuning, convolved with a kernel of width $u^{-1}$.  We plot several lineshapes in Fig.~\ref{f:lineshape}.

Essentially, we find that the steady-state behavior of the system indicates a maximum of population in the state opposite the initial state at a detuning $\delta = -\lambda \mean{\Az}$.  In addition, in the weak field limit, the line shape provides a sensitive measure of diagonal components of the bath's density matrix, so long as a calibration of $\Omega$ independent of the bath exists.

\subsection{Time-dependent behavior}
We now consider the actual, observable population and coherence oscillations induced by a driving (Rabi) field.  Looking at the explicit time dependence, we consider the envelope of oscillations given by $| \hat \zeta(t)|$.  The initial decay is quadratic, $(1-\mean{f})[1-(\gamma_{\Omega,0} t)^2/2]$, and when renormalized to give maximal contrast, it has a width 
\ba
\gamma_{\Omega,0} &=& \sqrt{-\frac{d^2}{dt^2} \Big| \frac{ \mean{\hat \zeta(t)} }{(1-\mean{\hat f})}\Big|_{t=0}} \\
&=& \frac{\Omega^2}{1-\mean{\hat f}} \times \\
 & &\sqrt{\Mean{\frac{1}{(\lambda \Az + \delta)^{2}+\Omega^2}}-\Mean{\frac{1}{\sqrt{(\lambda \Az + \delta)^2+\Omega^2}}}^2} . \nonumber
\ea
This suggests that measurements of such short time decay will measure second order moments of the bath operators.

While the short time behavior is consistent with Gaussian decay, as
would be expected from a non-Markovian, non-dissipative bath, 
the long time behavior is a power law.  

The integrals involved are solved in Appendix~\ref{s:math}; essentially we solve for $\mean{\hat \zeta(t)}$, and find
\beq
\mean{\zeta (t)} \approx \rho(-\delta) e^{-i \Omega t} \frac{\sqrt{2 \pi \Omega} e^{-i \theta(t)}}{\sqrt{\tau(t)}}. \label{e:finalRabiReal}
\eeq
where the time scale for phase shift, and phase shift angle, are given by
\ba
\tau(t) &=& \sqrt{t^2 + u^2} \\
\theta(t) &=& 1/2 \tan^{-1}\left[t / u\right] ,
\ea
and $u$ is the same as Eqn.~\ref{e:u}.

We note that this result is much more robust than
Eqn.~\ref{e:lineshapeReal}, due to the oscillatory terms in the
integrand canceling out behavior in the exponential tail.  This
solution encompasses the long time behavior ($t \gg \Omega^{-1}$), and
breaks down if $\rho_{sym}$ is singular for $\omega \ge \Omega$ or if
$u<0$~\cite{foot4}.
The lack of dependence of the
effective Rabi frequency on the detuning follows naturally from
situations where the density of states at $-\delta$, $\rho(-\delta)$
is sizeable: that portion of the density of states is resonant, and
the resonant behavior dominates.  When $u<0$, we may be in the
detuning dominated regime, and these results no longer hold.

\subsection{Discussion}
The long-time behavior, Eqn.~\ref{e:finalRabiReal}, may seem peculiar at first.  The power-law decay envelope comes from the portion of the bath density of states that is ``on-resonance'' with the oscillations.  In particular, for short times, Rabi oscillations are insensitive to detunings less than the Rabi frequency.  This window narrows for longer time, leading to the observed power law behavior.  More curiously, for times $t \gg u$, there is an overall phase shift of $\pi/4$ that is independent of many of the parameters of the bath.  This dynamic shift eludes a similar immediate qualitative description, arising from the continuity of the bath density of states and the pole induced by probing of the bath (by means of the applied Rabi field) that is insensitive to first derivatives of the bath, {\em i.e.}, bath variables come in only in their quadrature and higher moments.

It is crucial to distinguish these results from more standard inhomogeneous broadening results.  Unlike the case of Doppler broadening, for example, in this system the individual system detunings do not change as a function of time.  When the correlation function is short-lived, the behavior would instead follow directly from the well understood results describing inhomogenous broadening due to Doppler shifts of laser-induced Rabi oscillations.
To extend these results to describing inhomogeneous broadening in ensembles of self-assembled quantum dots, for example, would require considering the additional effect of inhomogeneity in the Rabi strength, $\Omega$, and it is unclear that the corresponding phase shift would behave in a similar manner to the case of a single Rabi frequency.  As such, that analysis is beyond the scope of this paper.

\begin{figure}
\begin{center}
\includegraphics[width=3.0in]{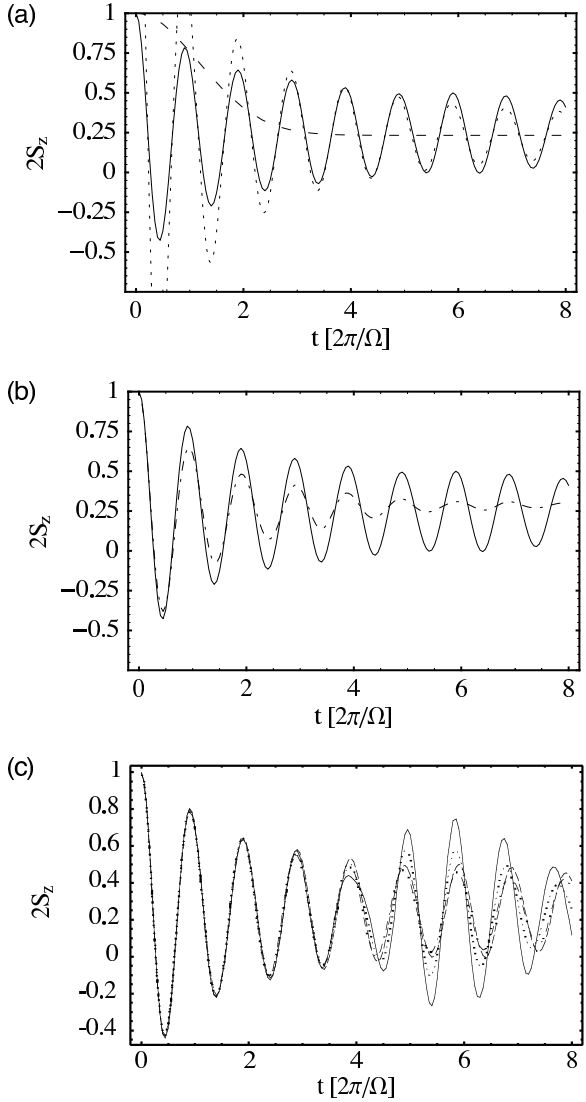}
\caption{
(a)
Comparison of short time and long time theoretical results to analytic results, with $\sigma_{\alpha} = \bar \alpha/2$, $\lambda = 1$, $N = 20$, and $\Omega = 2 \pi$.
The exact solution (solid line) compares well with the short time decay (dashed line) and the long time tail plus phase shift (dotted line).
(b)
Comparison of the same exact result with the case with $\gamma = 0.5$ (dotted line).
(c)
Fluctuations for $N=$6 (solid),10 (dashed), 14 (dashed-dotted), and 20 (dotted).
\label{f:rabi}
}
\end{center}
\end{figure}

To compare these analytical results to exact solutions, we solve the system for finite spin, where we exactly evaluate the trace for finite $N$.  In the homogeneous case, the Dicke picture of collective angular momentum allows us to go to large $N$, while the inhomogeneous case requires exponential operations for the exact value, but can be simulated through stochastic modeling of the trace function.  We compare several different values of inhomogeneity and bath size $N$ in Figure~\ref{f:rabi}, and show strong quantitative agreement between the short time and long time approximations with the numerical evolution.  We also show the convergence to full contrast for a fixed time and varying Rabi power.

The previous analysis assumed that the bath density of states could be taken as a Gaussian, which can be shown to reduce to the assumption of Gaussian noise~\cite{gardiner85}.  However, inclusion of non-Gaussian effects leads in the general cases to integrals without easy solution. 

The deviation from Gaussian noise for the fourth order at $T=\infty$ is
\beq
\frac{\mean{ \Az^4 } - 3 \mean{ \Az^2 }^2}{3 \mean{ \Az^2 }^2} \simeq
    - 2 / N .
\eeq
To investigate the role non-Gaussian effects play, simulations with finite numbers of spin were conducted; the thermodynamic limit recovers Gaussian statistics.  For small inhomogeneity, Fig.~\ref{f:rabi}c indicates that increasing spin number decreases the amplitude of fluctuations from the expected Rabi signal.  Increasing inhomogeneity leads to an apparent additional reduction of fluctuations, though quantifying this effect remains difficult.

\subsection{Inclusion of the Markovian environment}

For completeness, we now add the coupling of the qubit to an additional, Markovian environment, with a corresponding ``radiative'' decay to the ground state with a rate $\gamma$.  The optical Bloch equations 
must now be modified to include this decay, and exact solutions, while available, are cumbersome.  However, the case of FID, with $\Omega=0$, is immediately understandable.  In the Heisenberg picture, we write
\beq
\frac{d \Sp}{dt} = - \gamma \Sp + i \lambda \Az \Sp + \hat{\mathcal F}_+ ,
\eeq
where the stochastic input field $\hat{\mathcal F}_+$ has the standard Markovian kernel,
$\mean{ \hat{\mathcal F}_-(t)  \hat{\mathcal F}_+(t')} = 2 \gamma
\delta(t-t')$.  Solving this equation of motion exactly for
$\mean{\Sp(t=0)} = 1/2$ gives
\beq
\Phi_{FID,\gamma} = \frac{\mean{\Sp(t)}}{\mean{\Sp(0)}} = \Mean{ \exp(-i \lambda \Az t - \gamma t)}_{bath} = \Phi_{FID} e^{-\gamma t} .
\eeq
The Markovian environment results in {\em irreversible} exponential
decay, which would be unaffected by any spin-echo experiment.  For
long correlation time baths, spin-echo experiments provide a direct
measure of the Markovian component of the decay.

For the case with $\Omega \neq 0$, we use the analytical evolution
for solving the Kraus operator form for the Bloch vector,
$d \SV /dt = M \SV + \vec{u}$ 
and evaluating the expectation values of all functions of $\Az$ at
each time point. The results are cumbersome, and checked only
numerically. Qualitatively comparing the result, shown in
Figure~\ref{f:rabi}b, to the $\gamma=0$ case, we see that the crucial
features of phase shift and fast initial damping with long-lived
oscillations remain, but that the steady state population difference
is shifted, and the oscillations are eventually damped by the
exponential tail.  The corresponding lineshape is approximately
\beq
\mean{\hat f} = 1 - \Mean{\frac{\Omega \gamma}{[(\gamma/2)^2 + (\lambda \Az + \delta)^2 + \Omega^2]}}
\eeq

Finally, we now consider the effect decorrelation of $\Az$ plays in
the observed oscillations.  Unlike the case of free evolution,
$[H_0(t),H_0(t')] \neq 0$, leading to difficulties in analytical
evaluation of the propagator including decorrelation effects.  We can,
however, consider it partially by use of a Magnus expansion, in the
limit of weak to intermediate strength mesoscopic spin bath.

Formally, we transform $H_0$ to the interaction picture with respect
to $\Omega\Sx$, and with $\delta = 0$, giving
\beq
\tilde{H} = \lambda \Az [\cos(\Omega t) \Sz + \sin(\Omega t) \Sy]
\eeq
If we assume $\Az$ varies on time scales much longer than $1/\Omega$,
we can use a magnus expansion.  The expansion of this Hamiltonian at
peaks in the expected oscillations ($\tau = 2 \pi / \Omega$) yeilds
\beq
U_{eff}(\tau) = \exp[-i \tau (H_0 + H_1 + \ldots)]
\eeq
with
\ba
H_0 = 0 \\
H_1 = 2 (\lambda \Az)^2/\Omega \Sy\ .
\ea
We neglect higher order contributions by assuming that $\mean{\Az^2}^4 \lambda^4/\Omega^4 \ll 1$.
Thus, the envelope of oscillations should be determined by
\beq
U_{eff}(t) = \exp[-i \int_0^t \frac{2 (\lambda \Az(t'))^2}{\Omega} dt' \Sy]\ .
\eeq
where we have assumed that over a time $\tau$, $\Az(t)$ is fixed.  We
note that this form of quadratic noise has been investigated in
detail~\cite{makhlin04}.  The lowest order perturbative expansion
requires determining the spectral function
\begin{equation}
  \label{eq:S}
  S_{A^2}(\omega) = \int \frac{d\tau}{2 \pi} \mean{\Az(t+\tau)^2
    \Az(t)^2} e^{-i \omega \tau}\ .
\end{equation}
Assuming Gaussian statistics for $\Az$, we find
\begin{equation}
  \mean{Az(t)^2\Az(t')^2} = (\int d\omega S(\omega))^2 + 2 (\int
  d\omega S(\omega) e^{i \omega (t-t')})^2
\end{equation}

\begin{figure}
\includegraphics[width=3.0in]{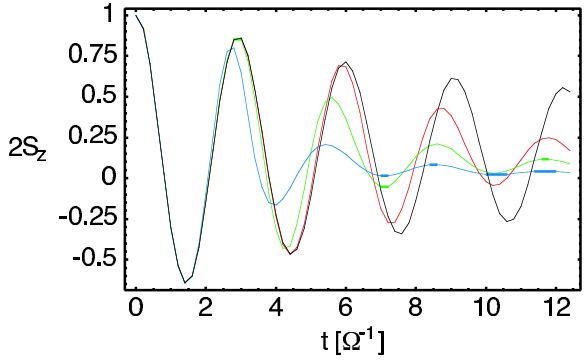}
\caption{
Expected value of $2 \Sz$ as a function of time for an initial state
$\Sz=1/2$, with driving and low frequency noise $\lambda = \Omega/2$.
Decorrelation rates of 0 (black), $\Omega/100$ (red), $\Omega/30$
(green) and $\Omega/10$ (blue) are shown.  It is apparent that
decorrelation leads to steady state values of zero, faster than
$1/\sqrt{t}$ decay, and reduced phase shift (from $\pi/4$).
\label{f:var}}
\end{figure}

The resulting decay of oscillations should be given by
\begin{equation}
  \exp(- \frac{4 \lambda^4}{\Omega^2} \int d\omega S_{A^2}(\omega)
  \frac{\sin^2(\omega t/2)}{(\omega/2)^2})
\end{equation}
Oscillations with low frequency noise for $S(\Omega)$ are shown in
Fig.~\ref{f:var}. 

We remark that for time independent $\Az$ and a Gaussian
density of states we can evaluate the expected envelope of
oscillations,
\beq
\mean{U_{eff}(t)} = (1+2 i t \rho''(0)/\rho(0) \Omega)^{-1/2}\ ,
\eeq
reproducing the $1/\sqrt{t}$ long-time tail found by more exact means.
Furthermore, as $\mean{\Az^2} \neq 0$ in general, a phase shift due to
small rotations around $\Sy$ after each Rabi oscillation should be
expected.

\section{Application to experimental systems}

We now consider the physical systems where the quasi-static bath
assumptions are valid.  One case is nuclear spins interacting with an
electron spin in a quantum dot.  There, the dominant internal bath
Hamiltonian is aligned with the bath-interaction, and $[\hat H_B, \Az
] = 0$ identically~\cite{taylor03, mehring76}.  Another system is
superconducting qubits, where charge-traps and other few-state
fluctuators play a similar role.  Finally, some NMR systems with one
probed spin (e.g., a Carbon-13) is coupled to many spins (e.g., nearby
hydrogens).

\subsection{Electron spin in quantum dots}

The hyperfine interaction between the electron and the nuclear spins
is given by
\beq
H_{QD} = \gamma_e B_0 \Sz + \lambda_{QD} \sum_{k} \alpha_k \SV \cdot \IV + \gamma_n B_0 \sum_k \Iz^k 
\eeq
where the $\alpha_k \propto |\psi(r_k)|^2$ are weighted by the
electron wavefunction's overlap with lattice site $k$.  For GaAs
quantum dots with $N$ nuclear spins, $\lambda_{QD} \simeq 207 {\rm
  ns}^{-1} / \sqrt{N}$~\cite{paget}, with the normalization condition
$\sum_k \alpha_k^2 = 1$.  Identifying $\gamma_e B_0 = \omega$ and
adding a time dependent ESR field in the $x$-axis the model
Hamiltonian is recovered, with $\lambda = \lambda_{QD}$.  In the case
of a large applied magnetic field ($\omega \gg \lambda_{QD}$), the
results heretofore derived hold.

While experiments investigating ESR and spin dephasing in single
quantum dots are ongoing, the contraints imposed by the quasi-static
limit make standard ESR technically difficult to achieve.  For small
quantum dots (e.g. single electron quantum dots), $\lambda_{QD}
\gtrsim 0.1$ ns$^{-1}$ ($N \simeq 10^5$).  If the decorrelation rate
$\Gamma$ is determined by dipolar-diffusion processes, it can be no
faster than 10 ms$^{-1}$, the linewidth of NMR in bulk GaAs, and the
quasi-static limit is appropriate.  For the quasi-static field to be
overcome, either many measurements must be taken in the correlation
time of the bath, or ESR field's Rabi frequency $\Omega$ must be much
larger than $0.1$ ns$^{-1}$.  This latter requirement may be quite
difficult for current experimental parameters. Active correction
sequences may lower this rate, by averaging through NMR pulses the
dipolar interaction~\cite{whh,mehring76}, making it possible to
perform accurate phase estimation within the enhanced correlation time
of the bath.

There is another important situation where the mesoscopic bath picture
holds: during an exchange gate~\cite{loss98,petta05}.  In particular, for two
tunnel coupled quantum dots, each in the single-electron regime, the
overall Hamiltonian is
\beq
H_{DQD} = H_{QD}^1 + H_{QD}^2 + J \SV^1 \cdot \SV^2
\eeq
where $J$ is the exchange interaction between the two dots. While the
$m_s=\pm1$ triplet states ($\ket{\upp \upp},\ket{\downn \downn}$) are
well separated by Zeeman energy, the $m_s=0$ states ($\ket{\upp
  \downn},\ket{\downn \upp}$) are nearly degenerate.  In the basis
$\{\ket{\upp \downn},\ket{\downn \upp}\}$, this two-level system is
described by the Hamiltonian
\beq
H_{m_s=0} = 
\left(
\begin{array}{cc}
- \lambda \Az^{(DQD)}/2 & \Omega/2 \\
\Omega/2 & \lambda \Az^{(DQD)}/2
\end{array}\right)
\eeq
where $\lambda \Az^{(DQD)} = \sum_{k,i=1,2} \lambda_{QD}^i (-)^{i} \alpha_k^i
\Iz^{k,i}$ is the bath parameter for this two-level system, and
$\Omega = J$.  This Hamiltonian maps exactly to Equation~\ref{e:hnot}.

Recently experiments demonstrated coherent oscillations driven by
controlled exchange interactions~\cite{petta05}.  The system was
prepared in an eigenstate of the mesoscopic spin bath interaction
(e.g., the $\ket{\upp \downn}$ state) and coherent oscillations
between $\ket{\upp \downn}$ and $\ket{\downn \upp}$ were driven by
applying a pulse of non-zero $\Omega$ for finite time.  These
experiments correspond exactly to the driven evolution examined in the
previous section.  As such, for small exchange values ($\Omega \simeq
\lambda$) we expect the exchange operations to exhibit the
characteristic power-law decay and phase shift found in this work.
Our approach agrees with system-specific theoretical predictions of
Ref.~\onlinecite{coish05}.

\subsection{Superconducting qubits}
The quasi-static assumptions also hold for bistable two-level
fluctuators, where the $\Iz$ eigenvalues denote which state is
populated, and the dominant interaction with a charge system is a
conditional capacitive energy, leading to a $\Sz \Iz$ interaction
between the two.  Furthermore, $1/f$-type
two-level fluctuators, considered the source of most Josephson
junction dephasing~\cite{kautz90,galperin91}, are quasi-static in the
high frequency regime, {\em i.e.}, above the bath cutoff.  Curiously,
the spin-boson model~\cite{caldeira83} does not fulfill this
criterion.
Other pseudo-spin systems, such as flux qubits, may or may not have
this structure, depending on the bath degrees of freedom.

The extension of our results to superconducting qubit experiments is
motivated by results characterized by low contrast, long-lived Rabi
oscillations, similar to those exhibited in our model.  For a system
with some minimum time of applied Rabi field (as is often the case in
these experiments) the initial, fast decay could be encompassed by
oscillations occurring within this minimum time, while the observable
oscillations correspond to the long-time tail.

Of the superconducting qubit experiments, we focus on
Ref.~\onlinecite{martinis02}, as 1/f noise is a dominant term for
similar devices in CW behavior.  In addition, there are observed
resonances with flucuators in the measurement
spectrum~\cite{simmonds04}, which suggests that coherent coupling to
these two-level systems may be possible. A model which includes
preparation and measurement errors is considered in
Appendix~\ref{s:super}; we note the results here. We find that in the
low power regime, our model explains the observed behavior, but at
higher powers incoherent heating due to driving microwaves play an
important role as well.

Fitting the parameters from Appendix~\ref{s:super} to the Rabi
oscillations shown in Ref.~\cite{martinis02} yield the values in
Table~\ref{t:expt}.  The results are consistent with bath domination
for low power ($\Omega < 0.5 {\rm ns}^{-1}$) and heating / preparation
error dominating at high power.  This gives the characteristic low
contrast oscillations observed, even though the static measurement
efficiency is quite high (order 75-85\%).  The resulting oscillations
are compared to Ref.~\onlinecite{martinis02} in Fig.~\ref{f:expt}.

\begin{table}
\caption{Fitted values (explained in Appendix~\ref{s:super}) for the
  model describing the experiments of Ref.~\onlinecite{martinis02}.  The values in parantesis are the 90\% confidence intervals of the fit.}
\label{t:expt}
\begin{ruledtabular}
\begin{tabular}{||c|c||} \hline
$M_{\upp \upp}$ & 0.75(4) \\
$M_{\downn \downn}$ & 1.00(3) \\
$\Gamma$ &0.10(4) ns$^{-1}$ \\
$\lambda$ & 0.27(1)  ns$^{-1}$ \\
\end{tabular}
\end{ruledtabular}
\end{table}

\begin{figure}
\includegraphics[width=3in]{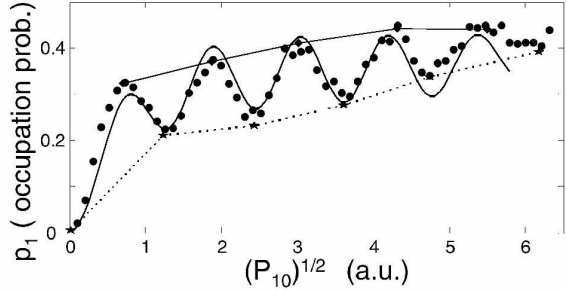}
\caption{Plot of oscillations reproduced from Ref.~\onlinecite{martinis02}
  (filled circles).  The probability of spin flip, $P_1 =
  \frac{1}{2}-\mean{\Sz(t)}$ is plotted versus Rabi frequency $\Omega
  \propto P_{10}^{1/2}$  a fixed Rabi evolution time of
  25 ns.  Overlaid for the given fit parameters are the expected
  values from our model (solid line).
\label{f:expt}}
\end{figure}

\subsection{Other systems}
Under certain conditions, NMR systems may be described by a
quasi-static mesoscopic spin bath .  For example, large molecules or
crystalline structures with long correlation times and coherent
dynamics within the molecule, where one or few spins are coupled to
many adjacent spins in a static configuration, may be described by our
model.  In contrast, liquid state NMR with small molecules has bath
characteristics dominated by transient couplings to other molecules,
and our model is inappropriate.

One example is in cross-polarization experiments on solid-state
ferrocene, where defocusing sequences are applied to generate a
Hamiltonian equivalent to Eqn.~\ref{e:hnot}~\cite{chattah03}.  By
direct observation of the cross-polarization signal, driven evolution
may be observed.  We remark that previous experiments, investigating
spin-bath coupling in free evolution~\cite{teklemariam03} will not
show this behavior.  System specific work~\cite{danieli05} is in general
agreement with our results.

\section{Methods to mitigate bath effects}
While the strongly coupled mesoscopic spin bath leads to dephasing
during free evolution and reduced contrast and a dynamical phase shift
during driven evolution, the long-time correlation of the bath allows
for correction.  We start by considering spin-echo-type passive error
correction, which can be implemented if a sufficiently strong Rabi
frequency for driving is available.  We then consider how entropy
transfer from the quasi-static bath to the environment (bath-cooling)
may be an effective alternative means of reducing the damping due to
the bath.  In essence, for specific bath models, it may be possible to
prepare coherences in the bath which effectively reduce the
uncertainties engendered by the bath.  The inherently quantum
mechanical nature of this leads us to consider this mechanism as
environmental squeezing.

\subsection{Passive error correction}
With strong Rabi power ($\Omega \gg \lambda$), high fidelity Rabi
oscillations are possible even in the presence of the mesoscopic spin
bath.  This is the power-broadened limit.  The error of such pulses scales as
$(\lambda/\Omega)^2$ and as such determining the quasi-static bath
strength for a given system will yield the required Rabi power to
overcome the effects of the bath through passive error correction.

In this limit, spin-echo techniques work well to greatly extend the
lifetime of oscillations, and passive error
correction~\cite{zanardi97} will be straightforward to implement.
Higher order pulse sequences than those considered in Section III may
lead to further improvement~\cite{fachi05}.  Other techniques, such as
quantum bang-bang~\cite{viola98}, are already implicit in this
analysis in the form of coherent averaging theory~\cite{mehring76}.

\subsection{Cooling the bath}

When experimental limitations of Rabi power or other experimental
difficulties make minimization of the coupling strength impossible,
the quasi-static bath remains a problem uncorrectable by passive error
correction.  Then, active cooling techniques may be a useful
alternative.  We now outline how, for the specific case of scalar
interactions with the bath, qubit manipulation may lead to efficient
entropy transfer from the mesoscopic spin bath to the qubit system.
By exploiting resonances between the bath and system, we may cool the
bath degrees of freedom.  Furthermore, in the quasi-static limit,
coherences developed between bath spins can lead to additional
improvements through a form of squeezing.

To illustrate cooling, the explicit case of an electron spin in a
quantum dot is illuminating.  For that system, the full spin-coupling
(including the terms previously neglected in the rotating wave
approximation) is a scalar interaction~\cite{taylor03},
\beq
\lambda \SV \cdot \AV = \lambda \SV \cdot \left( \sum_k \alpha_k \IV^k \right) .
\eeq
The dominant terms in the  internal Hamiltonian of the nuclear spins is
\beq
H_A = \sum_k \omega_k \Iz^k\ ,
\eeq
due to the Larmor precession of the nuclear spins in the external
magnetic field.  When the effective splitting of the qubit (electron
spin) is comparable to $\omega_k$, the electron spin comes into
resonance with the nuclear spins.  In essence, by tuning the system
near such a ``noise-resonance'', the bath can be driven in a
double-resonance manner (analogous to electron-nuclear double
resonance, ENDOR),
where the on-axis ($z-z$) coupling is offset by the changing frequency
of the applied Rabi field.  Before, the internal spin dynamics were
considered to be much slower, {\em i.e.}, $\omega_k \ll | \omega +
\lambda \mean{\Az} |$.  For double-resonance, this condition is no
longer satisfied, and the quasi-static approximations begin to break
down.

For this resonant regime, the reader is referred to previous work on
electron spin cooling~\cite{imamoglu03,taylor03b} for the details of
the process.  The benefits of cooling are outlined here.  In the fully
cooled state, the quasi-static bath acts entirely as an additional
detuning on the system, as $\mean{\Az}^2=\mean{\Az^2}$.  Near such
high polarization, the quasi-static bath effects do not supress Rabi
and Ramsey-type effects, as the bath strength parameter at high
temperature ($\lambda \sqrt{\mean{ \Az^2}}$) is greatly reduced at low
temperature.  Quantitatively, this improvement is given by the ratio
of $\gamma_{FID}$ (Eqn.~\ref{e:gammaFID}) at high and low
polarizations, which is $\eta = \sqrt{1/(1-P)}$ for a polarization $P$
of the bath (neglecting inhomogeneous corrections).  In terms of
experimental observables, it should result in an increase in the
observed contrast of Rabi oscillations and improved Ramsey signals as
if the bath strength parameter, $\lambda$, had been reduce by the
improvement factor $\eta$.  The improvements will be limited by
coherences developed within the bath, by the correlation time of the
quasi-static bath, which will limit the lifetime of the cooling, and
by other, non-local (and correspondingly difficult to cool) degrees of
freedom, implicit in the Markovian environment component of this
model.

Cooling could proceed along the following lines.  The system is
prepared in a $\ket{\downn}$ state by coupling to the Markovian
environment, {\em e.g.}, by moving to a strongly coupled region of parameter
space.  This is the current preparation method for
superconductor-based qubit designs; in quantum dots, the energy
difference between spin states due to zeeman splitting or exchange
interactions allows preparation by controlled coupling to a nearby
Fermi sea~\cite{petta05}.  Then a weak, negatively detuned Rabi field
is applied for a short time (order one Rabi flop).  This process is
repeated many times.  Including state measurement after the Rabi field
will allow measurement of the cooling efficiency, as the bath will
lead to additional flip to the $\ket{\upp}$ state in the far detuned
regime.  In some cases, where the natural frequencies of the bath
degrees of freedom are large, resonances between the bath and system
may be observed.  Tuning to the ``red'' side of these resonances will
also force cooling to occur.

As cooling is effected only through the logical basis states (the
qubit states of up and down), it is not necessarily efficient.  For a
simple scalar coupling ({\em e.g.} a hyperfine interaction), cooling
proceeds due to the action of the $\Am$ operator, where
\beq
\Am = \sum_k \alpha_k \hat{I}_-^k
\eeq
However, states $\ket{\mathcal D}$ with the property $\Am
\ket{\mathcal D} = 0$ cannot cool under this action.  This is similar
in practice to single mode cooling, and as such it does not greatly
improve the bath polarization or entropy~\cite{taylor03b} for
intermediate or larger bath sizes.  However, these states do have a
useful symmetry property, namely they are ``dark'' under the action of
the collective lowering operator.  Thus, the natural state of the
density matrix after many cycles of cooling is
\beq
\hat \rho = \sum_{\beta} \rho(\beta) \ket{{\mathcal D}(\beta)}
\eeq
where the sum is over other degeneracies
(see~\cite{arecchi72,taylor03b} for details).  We now indicate how
cooling limited to polarizations such that $\eta$ remains order unity
can still be sufficient to greatly reduce the dephasing induced by the
mesoscopic spin bath.

In particular, if this cooling can be applied along the axis of Rabi
oscillations (rather than the $z$ axis), the effects of such marginal
cooling are immediately apparent.  In particular, the component of
noise in the $x$ and $y$ axes are averaged by the bias, $\omega$,
while the $z$ axis is only averaged by the driving field, $\Omega$.
If the $z$ component of the bath noise can be reduced, the
corresponding Rabi and Ramsey signals will be improved.  In the case
of homogeneous coupling ($\alpha_k = 1/\sqrt{N}$) to the environment,
it can be shown immediately that cooling into dark states decreases
the quadrature in the non-cooled axes, {\em i.e.},
\beq
\mean{\Az^2} = 1/\sqrt{N}
\eeq
which is a $\sqrt{N}$ improvement over the uncooled state.  As the
other quadratures correspond to noise suppressed by the bias, the
effect on experiment is immediate and possible quite useful.  Such
Rabi-axis cooling can be achieved for nuclei in quantum dots by first
cooling along the natural ($z$) axis, followed by a NMR $\pi/2$-pulse
to rotate the nuclear coherences to be parallel to the driving field.

The coherences between spins in the bath have a limited lifetime,
determined by the decay of the correlation functions of the bath.
Therefore, for this process to be useful, the qubit-bath interaction
must be much faster than the bath decorrelation time, but this is
exactly the quasi-static approximation already made.  Before each run
of a Rabi oscillation-type experiment, a cooling sequence could be
applied, to keep the bath degrees of freedom.  The resulting
oscillations should reflect the ``cooled'' bath results, and as such
could provide a substantial improvement over uncooled systems.

\section{Conclusions}

The response of a two-level system (a qubit) with driving fields
coupled to a local, mesoscopic collection of finite-level systems and
with additional coupling to a larger, Markovian environment has
analyzed in this paper.  The natural simplifications of the
Hamiltonian lead to a quasi-static regime, wherein bath correlation
functions are long-lived when compared to experimental manipulation
and measurement time scales.  However, decorrelation of the bath and
dynamic thermalization processes lead to apparent decay when a
single-qubit is manipulated and measured repeatedly.  The large
variations of bias produced by the quasi-static bath can be difficult
to correct with limited Rabi frequency ranges accessible to
experiments.

The simple case of free induction decay demonstrates the role
increasing bath strength and non-Markovian effects have to play on the
short term, non-exponential decay, while finite number effects can
lead to revivals and fluctuations, distinctly mesoscopic effects.
This illustrates explicitly the non-Gaussian nature of the bath.

In the presence of weak fields, a line-shape distinctive to the
spin-bath model is discovered, though inclusion of additional
(Markovian) decoherence and dephasing recovers the expected Lorentzian
behavior.  As the driving field's strength is increased, Rabi
oscillations are possible, but for short times there may be a phase
shift of order $\pi/4$ and fast initial decay of the oscillation
envelope.  However, longer times, with only a power law decay going as
$t^{-1/2}$, will most likely be dominated by other, exponential decay
processes.  As a result, a quasi-static bath could lead to reduced
contrast of oscillations that may be consistent with experimental
observations of superconducting qubit devices.  In addition,
experiments with single- and double-quantum dots may probe this
quasi-static bath model more directly.

Finally, the role of passive error correction, quantum bang-bang, and
environment cooling have been considered, with each appropriate for
different ranges of parameters.  It may be that as the understanding
of quasi-static, local baths increases the means to mitigate their
effect will become more apparent.  This is a first and crucial step
towards that end.

The authors would like to thank A.~Imamoglu, D.~Loss, J.~M.~Martinis,
H.~Pastawski, A.~Sorensen, and C.~van der Wal for helpful discussions.
This work has been supported in part by NSF and ARO.

\appendix
\section{Integrals involved in driven evolution \label{s:math}}
We now solve the integrals necessary to evaluate $\mean{\hat f}$ and $\mean{\hat \zeta (t)}$ for a given bath density matrix, $\rho_{bath}$. Taking the continuum limit for this density matrix, we can evaluate a function $h(\lambda \Az)$ as 
\ba
{\rm Tr}[ h(\lambda \Az) \hat \rho_{bath} ] &=& \sum_{\Lambda} h(\Lambda) \rho_{\Lambda} \\
&\rightarrow & \int_{-\infty}^{\infty} d\Lambda h(\Lambda) \rho(\Lambda)
\ea

Accordingly, the long time population difference, given by $\mean{\hat f}$ has the form
\beq
 \mean{\hat f} = 1- \sum_{\Lambda} \rho_{\Lambda} \bra{\Lambda} \frac{1}{1 + (\frac{\delta + \lambda \Az}{\Omega})^2} \ket{\Lambda}
\eeq
where $\Lambda$ is the eigenvalue of $\lambda \Az$.  Going to the continuum limit, and assuming the power is weak (appropriate for lineshape measurements),
\beq
\mean{\hat f} = 1- \int_{-\infty}^{\infty} d\Lambda~\rho(\Lambda) \frac{\Omega^2}{(\Lambda + \delta)^2 + \Omega^2}
\eeq
Furthermore, the integral depends only upon $\tilde \omega = \sqrt{(\Lambda + \delta)^2 + \Omega^2}$, so we define the symmetric density,
\beq
\rho_{sym}(\omega) = \rho(-\delta - \sqrt{\omega^2 - \Omega^2}) + \rho(-\delta + \sqrt{\omega^2 - \Omega^2}) .
\eeq
Then,
\ba
\mean{\hat f} =  1- \int_{\Omega}^{\infty} d\omega \frac{\omega}{\sqrt{\omega^2 - \Omega^2}} \rho_{sym}(\omega) 
 \frac{\Omega^2}{\omega^2} \\
=1-  \Omega^{1/2} \int_0^{\infty} d\tilde \omega \frac{\rho_{sym}(\tilde \omega + \Omega)}{(\tilde \omega/ \Omega+1) \sqrt{(\tilde \omega / \Omega + 2)(\tilde \omega)}} , \label{e:intls}
\ea
While this integral can be solved for certain density of states, we note that for small frequencies ($\lesssim \Omega$) a relatively flat density of states can be approximated in a Taylor series, while for large frequencies, the $1/\tilde \omega^2$ behavior kills the higher frequencies components of the bath.  Expanding all of these, except for the pole at $\tilde \omega = 0$
the non-singular, non-oscillatory terms in Eqn.~\ref{e:int} near $\tilde \omega = 0$ are
\ba
 \frac{\rho_{sym}(\tilde \omega + \Omega) }{(\tilde \omega/ \Omega+1) \sqrt{\tilde \omega / \Omega + 2}} &=& \sqrt{2} \rho(-\delta) \Big[1- 
u \tilde \omega + ...\Big] \\
 &\approx & \sqrt{2} \rho(-\delta) \exp \left[
-u \tilde \omega \right] \label{e:linear}
\ea
where
the time scale is
\beq
u = \frac{5}{4 \Omega} -\frac{\Omega \rho''(-\delta)}{\rho(-\delta)} .
\eeq

Solving the integral yields
\beq
\mean{\hat f} \simeq 1 - \rho(-\delta) \sqrt{2 \pi \Omega/u} \label{e:lineshape}
\eeq

The oscillations at large $t$ can be evaluated in a manner similar to the lineshape.
\ba
\mean{\hat \zeta(t)} = \int_{-\infty}^{\infty} d\Lambda \rho(\Lambda) \frac{\Omega^2 \exp(-i \sqrt{(\Lambda + \delta)^2 + \Omega^2} t)}{(\Lambda + \delta)^2 + \Omega^2} \\
 = \int_{\Omega}^{\infty} d\omega \frac{\omega}{\sqrt{\omega^2 - \Omega^2}} \rho_{sym}(\omega) 
 \frac{\Omega^2 \exp(-i \omega t)}{\omega^2} \\
= \Omega^{1/2} e^{-i \Omega t} \int_0^{\infty} d\tilde \omega \frac{\rho_{sym}(\tilde \omega + \Omega) \exp(-i \tilde \omega t)}{(\tilde \omega/ \Omega+1) \sqrt{(\tilde \omega / \Omega + 2)(\tilde \omega)}} , \label{e:int}
\ea
where we have transformed once more to an offset variable, $\tilde
\omega = \omega - \Omega$.  For this final integral we use the
stationary phase approximation at long times, requiring that
$\rho_{sym}(\omega \ge \Omega)$ not be singular.

This linear term (Eqn.~\ref{e:linear}) determines the corrections to
the stationary phase integral.  We now define a time dependent angle
and an effective time,
\ba
\theta(t) &=& 1/2 \tan^{-1}\left[t / u\right] \\
\tau(t) &=& \sqrt{t^2 + u^2} .
\ea
Using these substitutions, we can solve for the stationary phase result exactly:
\ba
\mean{\zeta (t)} \approx \sqrt{2 \Omega} \rho(-\delta) e^{-i \Omega t} \int_0^{\infty} d\tilde \omega
\frac{e^{-i \tilde \omega t - \tilde \omega u}}{\sqrt{\tilde \omega}} \\
=  \rho(-\delta) e^{-i \Omega t} \frac{\sqrt{2 \pi \Omega} e^{-i \theta(t)}}{\sqrt{\tau(t)}}. \label{e:finalRabi}
\ea

\section{Comparison with superconducting qubits: model \label{s:super}}

We develop a theory describing superconducting qubit systems that
includes both the spin-bath model of this paper and incoherent heating
from microwave pulses used to implement driving.  To characterize the
experiment, a simple model involving several static parameters was
used. One is the preparation efficiency, $I$, which is the probability
of preparing ``down'' (the desired initial state) minus the
probability of preparing ``up'' (through bad preparation, probably due
to high temperature).  $I$ is set entirely by the microwave
temperature. Assuming that the only signficant source of heating is
the incident microwave field, Stefan's law gives $T \propto
\sqrt{\Omega}$.  Accordingly, $I = \tanh ( \sqrt{\Gamma / \Omega})$,
where $\Gamma^2$ is a measure of the cooling power.

In addition, there is a finite probability of measuring the wrong
state, given by the conditional probabilities, $M_{\upp,\upp}$ and
$M_{\downn,\downn}$.  The first is the probability of measuring up if
the system is in the up state at the time of measurement, and the
second is the same, but for down.  Assuming that $M_{\upp,\upp} +
M_{\upp,\downn} = 1$ and similarly for the $\downn$ case, the measured
signal is then
\beq
2 s_z = M_{\upp \upp} - M_{\downn \downn} + I_p S_z(\downn) (2 M_{\upp \upp} + 2 M_{\downn \downn} - 2)
\eeq
where $S_z(\downn)$ is the result given by the model of this paper for
an initial $\downn$ state (see Eqns.~\ref{e:lineshape} and
\ref{e:finalRabi}, and recall that $2 S_z = {\rm Re}[\mean{\hat f +
  \hat \zeta(t)}]$).

To include bath effects, a Gaussian density of states (appropriate if
the size of the bath is $\gtrsim 20$ effective spins) is assumed.
Then $\rho(\delta) = \exp[-\delta^2/(2 \lambda^2)] / \sqrt{2 \pi
  \lambda^2}$, where $\lambda$ is the bath strength parameter.  While
non-Gaussian effects, leading to mesoscopic fluctuations, are
important, their inclusion leads to too many highly correlated fitting
parameters.


\begin{thebibliography}{54}
\expandafter\ifx\csname natexlab\endcsname\relax\def\natexlab#1{#1}\fi
\expandafter\ifx\csname bibnamefont\endcsname\relax
  \def\bibnamefont#1{#1}\fi
\expandafter\ifx\csname bibfnamefont\endcsname\relax
  \def\bibfnamefont#1{#1}\fi
\expandafter\ifx\csname citenamefont\endcsname\relax
  \def\citenamefont#1{#1}\fi
\expandafter\ifx\csname url\endcsname\relax
  \def\url#1{\texttt{#1}}\fi
\expandafter\ifx\csname urlprefix\endcsname\relax\def\urlprefix{URL }\fi
\providecommand{\bibinfo}[2]{#2}
\providecommand{\eprint}[2][]{\url{#2}}

\bibitem[{\citenamefont{Loss and DiVincenzo}(1998)}]{loss98}
\bibinfo{author}{\bibfnamefont{D.}~\bibnamefont{Loss}} \bibnamefont{and}
  \bibinfo{author}{\bibfnamefont{D.}~\bibnamefont{DiVincenzo}},
  \bibinfo{journal}{Phys. Rev. A} \textbf{\bibinfo{volume}{57}},
  \bibinfo{pages}{120} (\bibinfo{year}{1998}).

\bibitem[{\citenamefont{Imamoglu et~al.}(1999)\citenamefont{Imamoglu,
  Awschalom, Burkard, DiVincenzo, Loss, Sherwin, and Small}}]{imamoglu99}
\bibinfo{author}{\bibfnamefont{A.}~\bibnamefont{Imamoglu}},
  \bibinfo{author}{\bibfnamefont{D.~D.} \bibnamefont{Awschalom}},
  \bibinfo{author}{\bibfnamefont{G.}~\bibnamefont{Burkard}},
  \bibinfo{author}{\bibfnamefont{D.~P.} \bibnamefont{DiVincenzo}},
  \bibinfo{author}{\bibfnamefont{D.}~\bibnamefont{Loss}},
  \bibinfo{author}{\bibfnamefont{M.}~\bibnamefont{Sherwin}}, \bibnamefont{and}
  \bibinfo{author}{\bibfnamefont{A.}~\bibnamefont{Small}},
  \bibinfo{journal}{Phys. Rev. Lett.} \textbf{\bibinfo{volume}{83}},
  \bibinfo{pages}{4204} (\bibinfo{year}{1999}).

\bibitem[{\citenamefont{Merkulov et~al.}(2002)\citenamefont{Merkulov, Efros,
  and Rosen}}]{merkulov02}
\bibinfo{author}{\bibfnamefont{I.~A.} \bibnamefont{Merkulov}},
  \bibinfo{author}{\bibfnamefont{A.~L.} \bibnamefont{Efros}}, \bibnamefont{and}
  \bibinfo{author}{\bibfnamefont{M.}~\bibnamefont{Rosen}},
  \bibinfo{journal}{Phys. Rev. B} \textbf{\bibinfo{volume}{65}},
  \bibinfo{pages}{205309} (\bibinfo{year}{2002}).

\bibitem[{\citenamefont{{Khaetskii} et~al.}(2002)\citenamefont{{Khaetskii},
  {Loss}, and {Glazman}}}]{khaetskii02}
\bibinfo{author}{\bibfnamefont{A.~V.} \bibnamefont{{Khaetskii}}},
  \bibinfo{author}{\bibfnamefont{D.}~\bibnamefont{{Loss}}}, \bibnamefont{and}
  \bibinfo{author}{\bibfnamefont{L.}~\bibnamefont{{Glazman}}},
  \bibinfo{journal}{Phys. Rev. Lett.} \textbf{\bibinfo{volume}{88}},
  \bibinfo{pages}{186802} (\bibinfo{year}{2002}),
  \urlprefix\url{http://publish.aps.org/abstract/prl/v88/p186802}.

\bibitem[{\citenamefont{Taylor et~al.}(2003{\natexlab{a}})\citenamefont{Taylor,
  Marcus, and Lukin}}]{taylor03}
\bibinfo{author}{\bibfnamefont{J.~M.} \bibnamefont{Taylor}},
  \bibinfo{author}{\bibfnamefont{C.~M.} \bibnamefont{Marcus}},
  \bibnamefont{and} \bibinfo{author}{\bibfnamefont{M.~D.} \bibnamefont{Lukin}},
  \bibinfo{journal}{Phys. Rev. Lett.} \textbf{\bibinfo{volume}{90}},
  \bibinfo{pages}{206803} (\bibinfo{year}{2003}{\natexlab{a}}).

\bibitem[{\citenamefont{Johnson et~al.}(2005)\citenamefont{Johnson, Petta,
  Taylor, Lukin, Marcus, Hanson, and Gossard}}]{johnson05}
\bibinfo{author}{\bibfnamefont{A.~C.} \bibnamefont{Johnson}},
  \bibinfo{author}{\bibfnamefont{J.}~\bibnamefont{Petta}},
  \bibinfo{author}{\bibfnamefont{J.~M.} \bibnamefont{Taylor}},
  \bibinfo{author}{\bibfnamefont{M.~D.} \bibnamefont{Lukin}},
  \bibinfo{author}{\bibfnamefont{C.~M.} \bibnamefont{Marcus}},
  \bibinfo{author}{\bibfnamefont{M.~P.} \bibnamefont{Hanson}},
  \bibnamefont{and} \bibinfo{author}{\bibfnamefont{A.~C.}
  \bibnamefont{Gossard}}, \bibinfo{journal}{Nature}
  \textbf{\bibinfo{volume}{435}}, \bibinfo{pages}{925} (\bibinfo{year}{2005}).

\bibitem[{\citenamefont{Koppens et~al.}(2005)\citenamefont{Koppens, Folk,
  Elzerman, Hanson, Willems~van Beveren, Vink, Tranitz, Wegscheider,
  Kouwenhoven, and Vandersypen}}]{koppens05}
\bibinfo{author}{\bibfnamefont{F.~H.~L.} \bibnamefont{Koppens}},
  \bibinfo{author}{\bibfnamefont{J.~A.} \bibnamefont{Folk}},
  \bibinfo{author}{\bibfnamefont{J.~M.} \bibnamefont{Elzerman}},
  \bibinfo{author}{\bibfnamefont{R.}~\bibnamefont{Hanson}},
  \bibinfo{author}{\bibfnamefont{L.~H.} \bibnamefont{Willems~van Beveren}},
  \bibinfo{author}{\bibfnamefont{I.~T.} \bibnamefont{Vink}},
  \bibinfo{author}{\bibfnamefont{H.-P.} \bibnamefont{Tranitz}},
  \bibinfo{author}{\bibfnamefont{W.}~\bibnamefont{Wegscheider}},
  \bibinfo{author}{\bibfnamefont{L.~P.} \bibnamefont{Kouwenhoven}},
  \bibnamefont{and} \bibinfo{author}{\bibfnamefont{L.~M.~K.}
  \bibnamefont{Vandersypen}}, \bibinfo{journal}{Science} p.
  \bibinfo{pages}{1113719} (\bibinfo{year}{2005}),
  \urlprefix\url{http://www.sciencemag.org/cgi/content/abstract/1113719v2}.

\bibitem[{\citenamefont{Petta et~al.}(2005)\citenamefont{Petta, Johnson,
  Taylor, Laird, Yacoby, Lukin, and Marcus}}]{petta05}
\bibinfo{author}{\bibfnamefont{J.~R.} \bibnamefont{Petta}},
  \bibinfo{author}{\bibfnamefont{A.~C.} \bibnamefont{Johnson}},
  \bibinfo{author}{\bibfnamefont{J.~M.} \bibnamefont{Taylor}},
  \bibinfo{author}{\bibfnamefont{E.}~\bibnamefont{Laird}},
  \bibinfo{author}{\bibfnamefont{A.}~\bibnamefont{Yacoby}},
  \bibinfo{author}{\bibfnamefont{M.~D.} \bibnamefont{Lukin}}, \bibnamefont{and}
  \bibinfo{author}{\bibfnamefont{C.~M.} \bibnamefont{Marcus}},
  \bibinfo{journal}{Science} \textbf{\bibinfo{volume}{309}},
  \bibinfo{pages}{2180} (\bibinfo{year}{2005}).

\bibitem[{\citenamefont{de~Sousa and {Das
  Sarma}}(2003{\natexlab{a}})}]{desousa03b}
\bibinfo{author}{\bibfnamefont{R.}~\bibnamefont{de~Sousa}} \bibnamefont{and}
  \bibinfo{author}{\bibfnamefont{S.}~\bibnamefont{{Das Sarma}}},
  \bibinfo{journal}{Phys. Rev. B} \textbf{\bibinfo{volume}{67}},
  \bibinfo{pages}{033301} (\bibinfo{year}{2003}{\natexlab{a}}).

\bibitem[{\citenamefont{Coish and Loss}(2004)}]{coish04}
\bibinfo{author}{\bibfnamefont{W.~A.} \bibnamefont{Coish}} \bibnamefont{and}
  \bibinfo{author}{\bibfnamefont{D.}~\bibnamefont{Loss}},
  \bibinfo{journal}{Phys. Rev. B} \textbf{\bibinfo{volume}{70}},
  \bibinfo{pages}{195340} (\bibinfo{year}{2004}).

\bibitem[{\citenamefont{Elzerman et~al.}(2004)\citenamefont{Elzerman, Hanson,
  van Beveren, Witkamp, Vandersypen, and Kouwenhoven}}]{elzerman04}
\bibinfo{author}{\bibfnamefont{J.~M.} \bibnamefont{Elzerman}},
  \bibinfo{author}{\bibfnamefont{R.}~\bibnamefont{Hanson}},
  \bibinfo{author}{\bibfnamefont{L.~H.~W.} \bibnamefont{van Beveren}},
  \bibinfo{author}{\bibfnamefont{B.}~\bibnamefont{Witkamp}},
  \bibinfo{author}{\bibfnamefont{L.~M.~K.} \bibnamefont{Vandersypen}},
  \bibnamefont{and} \bibinfo{author}{\bibfnamefont{L.~P.}
  \bibnamefont{Kouwenhoven}}, \bibinfo{journal}{Nature}
  \textbf{\bibinfo{volume}{430}}, \bibinfo{pages}{431} (\bibinfo{year}{2004}).

\bibitem[{\citenamefont{Fujisawa et~al.}(2001)\citenamefont{Fujisawa, Tokura,
  and Hirayama}}]{fujisawa01}
\bibinfo{author}{\bibfnamefont{T.}~\bibnamefont{Fujisawa}},
  \bibinfo{author}{\bibfnamefont{Y.}~\bibnamefont{Tokura}}, \bibnamefont{and}
  \bibinfo{author}{\bibfnamefont{Y.}~\bibnamefont{Hirayama}},
  \bibinfo{journal}{Phys. Rev. B. (Rapid Comm.)} \textbf{\bibinfo{volume}{63}},
  \bibinfo{pages}{081304} (\bibinfo{year}{2001}).

\bibitem[{\citenamefont{Golovach et~al.}(2004)\citenamefont{Golovach,
  Khaetskii, and Loss}}]{golovach04}
\bibinfo{author}{\bibfnamefont{V.~N.} \bibnamefont{Golovach}},
  \bibinfo{author}{\bibfnamefont{A.}~\bibnamefont{Khaetskii}},
  \bibnamefont{and} \bibinfo{author}{\bibfnamefont{D.}~\bibnamefont{Loss}},
  \bibinfo{journal}{Phys. Rev. Lett.} \textbf{\bibinfo{volume}{93}},
  \bibinfo{pages}{016601} (\bibinfo{year}{2004}).

\bibitem[{\citenamefont{Hanson et~al.}(2003)\citenamefont{Hanson, Witkamp,
  Vandersypen, {van Beveren}, Elzerman, and Kouwenhoven}}]{hanson03}
\bibinfo{author}{\bibfnamefont{R.}~\bibnamefont{Hanson}},
  \bibinfo{author}{\bibfnamefont{B.}~\bibnamefont{Witkamp}},
  \bibinfo{author}{\bibfnamefont{L.~M.~K.} \bibnamefont{Vandersypen}},
  \bibinfo{author}{\bibfnamefont{L.~H.~W.} \bibnamefont{{van Beveren}}},
  \bibinfo{author}{\bibfnamefont{J.~M.} \bibnamefont{Elzerman}},
  \bibnamefont{and} \bibinfo{author}{\bibfnamefont{L.~P.}
  \bibnamefont{Kouwenhoven}}, \bibinfo{journal}{Phys. Rev. Lett.}
  \textbf{\bibinfo{volume}{91}}, \bibinfo{pages}{196802}
  (\bibinfo{year}{2003}).

\bibitem[{\citenamefont{Vion et~al.}(2002)\citenamefont{Vion, Aassime, Cottet,
  Joyez, Pothier, Urbina, Esteve, and Devorett}}]{vion02}
\bibinfo{author}{\bibfnamefont{D.}~\bibnamefont{Vion}},
  \bibinfo{author}{\bibfnamefont{A.}~\bibnamefont{Aassime}},
  \bibinfo{author}{\bibfnamefont{A.}~\bibnamefont{Cottet}},
  \bibinfo{author}{\bibfnamefont{P.}~\bibnamefont{Joyez}},
  \bibinfo{author}{\bibfnamefont{H.}~\bibnamefont{Pothier}},
  \bibinfo{author}{\bibfnamefont{C.}~\bibnamefont{Urbina}},
  \bibinfo{author}{\bibfnamefont{D.}~\bibnamefont{Esteve}}, \bibnamefont{and}
  \bibinfo{author}{\bibfnamefont{M.}~\bibnamefont{Devorett}},
  \bibinfo{journal}{Science} \textbf{\bibinfo{volume}{296}},
  \bibinfo{pages}{886} (\bibinfo{year}{2002}).

\bibitem[{\citenamefont{Pashkin et~al.}(2003)\citenamefont{Pashkin, Yamamoto,
  Astafiev, Nakamura, Averin, and Tsai}}]{pashkin03}
\bibinfo{author}{\bibfnamefont{Y.~A.} \bibnamefont{Pashkin}},
  \bibinfo{author}{\bibfnamefont{T.}~\bibnamefont{Yamamoto}},
  \bibinfo{author}{\bibfnamefont{O.}~\bibnamefont{Astafiev}},
  \bibinfo{author}{\bibfnamefont{Y.}~\bibnamefont{Nakamura}},
  \bibinfo{author}{\bibfnamefont{D.}~\bibnamefont{Averin}}, \bibnamefont{and}
  \bibinfo{author}{\bibfnamefont{J.}~\bibnamefont{Tsai}},
  \bibinfo{journal}{Nature} \textbf{\bibinfo{volume}{421}},
  \bibinfo{pages}{823} (\bibinfo{year}{2003}).

\bibitem[{\citenamefont{Chiorescu et~al.}(2003)\citenamefont{Chiorescu,
  Nakamura, Harmans, and Mooij}}]{chiorescu03}
\bibinfo{author}{\bibfnamefont{I.}~\bibnamefont{Chiorescu}},
  \bibinfo{author}{\bibfnamefont{Y.}~\bibnamefont{Nakamura}},
  \bibinfo{author}{\bibfnamefont{C.}~\bibnamefont{Harmans}}, \bibnamefont{and}
  \bibinfo{author}{\bibfnamefont{J.}~\bibnamefont{Mooij}},
  \bibinfo{journal}{Science} \textbf{\bibinfo{volume}{299}},
  \bibinfo{pages}{1869} (\bibinfo{year}{2003}).

\bibitem[{\citenamefont{Martinis et~al.}(2002)\citenamefont{Martinis, Nam,
  Aumentado, and Urbina}}]{martinis02}
\bibinfo{author}{\bibfnamefont{J.}~\bibnamefont{Martinis}},
  \bibinfo{author}{\bibfnamefont{S.}~\bibnamefont{Nam}},
  \bibinfo{author}{\bibfnamefont{J.}~\bibnamefont{Aumentado}},
  \bibnamefont{and} \bibinfo{author}{\bibfnamefont{C.}~\bibnamefont{Urbina}},
  \bibinfo{journal}{Phys. Rev. Lett.} \textbf{\bibinfo{volume}{89}},
  \bibinfo{pages}{117901} (\bibinfo{year}{2002}).

\bibitem[{\citenamefont{Simmonds et~al.}(2004)\citenamefont{Simmonds, Lang,
  Hite, Nam, Pappas, and Martinis}}]{simmonds04}
\bibinfo{author}{\bibfnamefont{R.}~\bibnamefont{Simmonds}},
  \bibinfo{author}{\bibfnamefont{K.~M.} \bibnamefont{Lang}},
  \bibinfo{author}{\bibfnamefont{D.~A.} \bibnamefont{Hite}},
  \bibinfo{author}{\bibfnamefont{S.}~\bibnamefont{Nam}},
  \bibinfo{author}{\bibfnamefont{D.~P.} \bibnamefont{Pappas}},
  \bibnamefont{and} \bibinfo{author}{\bibfnamefont{J.~M.}
  \bibnamefont{Martinis}}, \bibinfo{journal}{Phys. Rev. Lett.}
  \textbf{\bibinfo{volume}{93}}, \bibinfo{pages}{077003}
  (\bibinfo{year}{2004}).

\bibitem[{\citenamefont{Makhlin and Shnirman}(2004)}]{makhlin04}
\bibinfo{author}{\bibfnamefont{Y.}~\bibnamefont{Makhlin}} \bibnamefont{and}
  \bibinfo{author}{\bibfnamefont{A.}~\bibnamefont{Shnirman}},
  \bibinfo{journal}{Phys. Rev. Lett.} \textbf{\bibinfo{volume}{92}},
  \bibinfo{pages}{178301} (\bibinfo{year}{2004}).

\bibitem[{\citenamefont{Falci et~al.}(2005)\citenamefont{Falci, D'Arrigo,
  Mastellone, and Paladino}}]{falci05}
\bibinfo{author}{\bibfnamefont{G.}~\bibnamefont{Falci}},
  \bibinfo{author}{\bibfnamefont{A.}~\bibnamefont{D'Arrigo}},
  \bibinfo{author}{\bibfnamefont{A.}~\bibnamefont{Mastellone}},
  \bibnamefont{and} \bibinfo{author}{\bibfnamefont{E.}~\bibnamefont{Paladino}},
  \bibinfo{journal}{Phys. Rev. Lett.} \textbf{\bibinfo{volume}{94}},
  \bibinfo{pages}{167002} (\bibinfo{year}{2005}).

\bibitem[{\citenamefont{Stamp}(2003)}]{stamp03}
\bibinfo{author}{\bibfnamefont{P.}~\bibnamefont{Stamp}},
  \emph{\bibinfo{title}{The Physics of Communication}} (\bibinfo{publisher}{New
  Jersey: World Scientific}, \bibinfo{year}{2003}), chap.~\bibinfo{chapter}{3},
  pp. \bibinfo{pages}{39--82}.

\bibitem[{\citenamefont{Taylor et~al.}(2003{\natexlab{b}})\citenamefont{Taylor,
  Imamoglu, and Lukin}}]{taylor03b}
\bibinfo{author}{\bibfnamefont{J.~M.} \bibnamefont{Taylor}},
  \bibinfo{author}{\bibfnamefont{A.}~\bibnamefont{Imamoglu}}, \bibnamefont{and}
  \bibinfo{author}{\bibfnamefont{M.~D.} \bibnamefont{Lukin}},
  \bibinfo{journal}{Phys. Rev. Lett.} \textbf{\bibinfo{volume}{91}},
  \bibinfo{pages}{246802} (\bibinfo{year}{2003}{\natexlab{b}}).

\bibitem[{\citenamefont{Weissman}(1988)}]{weissman88}
\bibinfo{author}{\bibfnamefont{M.~B.} \bibnamefont{Weissman}},
  \bibinfo{journal}{Rev. Mod. Phys.} \textbf{\bibinfo{volume}{60}},
  \bibinfo{pages}{537} (\bibinfo{year}{1988}).

\bibitem[{\citenamefont{Klauser et~al.}(2005)\citenamefont{Klauser, Coish, and
  Loss}}]{klauser05}
\bibinfo{author}{\bibfnamefont{D.}~\bibnamefont{Klauser}},
  \bibinfo{author}{\bibfnamefont{W.~A.} \bibnamefont{Coish}}, \bibnamefont{and}
  \bibinfo{author}{\bibfnamefont{D.}~\bibnamefont{Loss}},
  \bibinfo{journal}{e-print: cond-mat/0510177}  (\bibinfo{year}{2005}).

\bibitem[{\citenamefont{Coish and Loss}(2005)}]{coish05}
\bibinfo{author}{\bibfnamefont{W.~A.} \bibnamefont{Coish}} \bibnamefont{and}
  \bibinfo{author}{\bibfnamefont{D.}~\bibnamefont{Loss}},
  \bibinfo{journal}{Phys. Rev. B} \textbf{\bibinfo{volume}{72}},
  \bibinfo{pages}{125337} (\bibinfo{year}{2005}).

\bibitem[{\citenamefont{Hu and Sarma}(2005)}]{hu05}
\bibinfo{author}{\bibfnamefont{X.}~\bibnamefont{Hu}} \bibnamefont{and}
  \bibinfo{author}{\bibfnamefont{S.~D.} \bibnamefont{Sarma}},
  \bibinfo{journal}{e-print: cond-mat/0507725}  (\bibinfo{year}{2005}).

\bibitem[{\citenamefont{Zurek}(1981)}]{zurek81}
\bibinfo{author}{\bibfnamefont{W.~H.} \bibnamefont{Zurek}},
  \bibinfo{journal}{Phys. Rev. D} \textbf{\bibinfo{volume}{24}},
  \bibinfo{pages}{1516} (\bibinfo{year}{1981}).

\bibitem[{\citenamefont{Prokof'ev and Stamp}(2000)}]{prokofev00}
\bibinfo{author}{\bibfnamefont{N.~V.} \bibnamefont{Prokof'ev}}
  \bibnamefont{and} \bibinfo{author}{\bibfnamefont{P.~C.~E.}
  \bibnamefont{Stamp}}, \bibinfo{journal}{Reports on Progress in Physics}
  \textbf{\bibinfo{volume}{63}}, \bibinfo{pages}{669} (\bibinfo{year}{2000}).

\bibitem[{\citenamefont{Rose and Smirnov}(2001)}]{rose01}
\bibinfo{author}{\bibfnamefont{G.}~\bibnamefont{Rose}} \bibnamefont{and}
  \bibinfo{author}{\bibfnamefont{A.~Y.} \bibnamefont{Smirnov}},
  \bibinfo{journal}{J. Phys.: Cond. Mat.} \textbf{\bibinfo{volume}{13}},
  \bibinfo{pages}{11027} (\bibinfo{year}{2001}).

\bibitem[{\citenamefont{Zanardi and Rasetti}(1997)}]{zanardi97}
\bibinfo{author}{\bibfnamefont{P.}~\bibnamefont{Zanardi}} \bibnamefont{and}
  \bibinfo{author}{\bibfnamefont{M.}~\bibnamefont{Rasetti}},
  \bibinfo{journal}{Phys. Rev. Lett.} \textbf{\bibinfo{volume}{79}},
  \bibinfo{pages}{3306} (\bibinfo{year}{1997}).

\bibitem[{\citenamefont{Viola and Lloyd}(1998)}]{viola98}
\bibinfo{author}{\bibfnamefont{L.}~\bibnamefont{Viola}} \bibnamefont{and}
  \bibinfo{author}{\bibfnamefont{S.}~\bibnamefont{Lloyd}},
  \bibinfo{journal}{Phys. Rev. A} \textbf{\bibinfo{volume}{58}},
  \bibinfo{pages}{2733} (\bibinfo{year}{1998}).

\bibitem{foot1}
The breakdown of the two-level
  approximation in superconductor-based qubit designs has already been
  explored in great detail (Burkard {\em et al.} Phys. Rev
  B. \textbf{69}, 064503 (2004)) and we instead focus on
  other sources of error due to local spins, charge traps, etc.


\bibitem[{\citenamefont{Feynman and Vernon}(1963)}]{feynman63}
\bibinfo{author}{\bibfnamefont{R.~P.} \bibnamefont{Feynman}} \bibnamefont{and}
  \bibinfo{author}{\bibfnamefont{F.~L.} \bibnamefont{Vernon}},
  \bibinfo{journal}{Annals of Physics} \textbf{\bibinfo{volume}{24}},
  \bibinfo{pages}{118} (\bibinfo{year}{1963}).

\bibitem[{\citenamefont{Magnus}(1954)}]{magnus54}
\bibinfo{author}{\bibfnamefont{W.}~\bibnamefont{Magnus}},
  \bibinfo{journal}{Communications of Pure and Applied Mathematics}
  \textbf{\bibinfo{volume}{7}}, \bibinfo{pages}{649} (\bibinfo{year}{1954}).

\bibitem[{\citenamefont{Cottet and {\em et al.}}(2001)}]{cottet01}
\bibinfo{author}{\bibfnamefont{A.}~\bibnamefont{Cottet}} \bibnamefont{and}
  \bibinfo{author}{\bibnamefont{{\em et al.}}},
  \emph{\bibinfo{title}{Macroscopic Quantum Coherence and Quantum Computing}}
  (\bibinfo{publisher}{Kluwer/Plenum}, \bibinfo{address}{New York},
  \bibinfo{year}{2001}), p. \bibinfo{pages}{111}.

\bibitem[{\citenamefont{Giedke et~al.}(2005)\citenamefont{Giedke, Taylor,
  D'Alessandro, Lukin, and Imamoglu}}]{giedke05}
\bibinfo{author}{\bibfnamefont{G.}~\bibnamefont{Giedke}},
  \bibinfo{author}{\bibfnamefont{J.~M.} \bibnamefont{Taylor}},
  \bibinfo{author}{\bibfnamefont{D.}~\bibnamefont{D'Alessandro}},
  \bibinfo{author}{\bibfnamefont{M.~D.} \bibnamefont{Lukin}}, \bibnamefont{and}
  \bibinfo{author}{\bibfnamefont{A.}~\bibnamefont{Imamoglu}},
  \bibinfo{journal}{e-print: quant-ph/0508144}  (\bibinfo{year}{2005}).

\bibitem[{\citenamefont{Taylor et~al.}(2005)\citenamefont{Taylor, Petta,
  Johnson, Yacoby, Marcus, and Lukin}}]{taylor05prb}
\bibinfo{author}{\bibfnamefont{J.~M.} \bibnamefont{Taylor}},
  \bibinfo{author}{\bibfnamefont{J.}~\bibnamefont{Petta}},
  \bibinfo{author}{\bibfnamefont{A.~C.} \bibnamefont{Johnson}},
  \bibinfo{author}{\bibfnamefont{A.}~\bibnamefont{Yacoby}},
  \bibinfo{author}{\bibfnamefont{C.~M.} \bibnamefont{Marcus}},
  \bibnamefont{and} \bibinfo{author}{\bibfnamefont{M.~D.} \bibnamefont{Lukin}},
  \bibinfo{journal}{(in preparation)}  (\bibinfo{year}{2005}).

\bibitem[{\citenamefont{de~Sousa and {Das
  Sarma}}(2003{\natexlab{b}})}]{desousa03}
\bibinfo{author}{\bibfnamefont{R.}~\bibnamefont{de~Sousa}} \bibnamefont{and}
  \bibinfo{author}{\bibfnamefont{S.}~\bibnamefont{{Das Sarma}}},
  \bibinfo{journal}{Phys. Rev. B} \textbf{\bibinfo{volume}{68}},
  \bibinfo{pages}{115322} (\bibinfo{year}{2003}{\natexlab{b}}).

\bibitem[{\citenamefont{Yao et~al.}(2005)\citenamefont{Yao, Liu, and
  Sham}}]{yao05}
\bibinfo{author}{\bibfnamefont{W.}~\bibnamefont{Yao}},
  \bibinfo{author}{\bibfnamefont{R.-B.} \bibnamefont{Liu}}, \bibnamefont{and}
  \bibinfo{author}{\bibfnamefont{L.~J.} \bibnamefont{Sham}},
  \bibinfo{journal}{e-print: cond-mat/0508441}  (\bibinfo{year}{2005}).

\bibitem[{\citenamefont{Deng and Hu}(2005)}]{deng03}
\bibinfo{author}{\bibfnamefont{C.}~\bibnamefont{Deng}} \bibnamefont{and}
  \bibinfo{author}{\bibfnamefont{X.}~\bibnamefont{Hu}}, \bibinfo{journal}{Phys.
  Rev. B} \textbf{\bibinfo{volume}{72}}, \bibinfo{pages}{165333}
  (\bibinfo{year}{2005}).

\bibitem[{\citenamefont{Paget}(1982)}]{paget83}
\bibinfo{author}{\bibfnamefont{D.}~\bibnamefont{Paget}},
  \bibinfo{journal}{Phys. Rev. B} \textbf{\bibinfo{volume}{25}},
  \bibinfo{pages}{4444} (\bibinfo{year}{1982}).

\bibitem[{\citenamefont{Teklemariam et~al.}(2003)\citenamefont{Teklemariam,
  Fortunato, Lopez, Emerson, Paz, Havel, and Cory}}]{teklemariam03}
\bibinfo{author}{\bibfnamefont{G.}~\bibnamefont{Teklemariam}},
  \bibinfo{author}{\bibfnamefont{E.~M.} \bibnamefont{Fortunato}},
  \bibinfo{author}{\bibfnamefont{C.~C.} \bibnamefont{Lopez}},
  \bibinfo{author}{\bibfnamefont{J.}~\bibnamefont{Emerson}},
  \bibinfo{author}{\bibfnamefont{J.~P.} \bibnamefont{Paz}},
  \bibinfo{author}{\bibfnamefont{T.~F.} \bibnamefont{Havel}}, \bibnamefont{and}
  \bibinfo{author}{\bibfnamefont{D.~G.} \bibnamefont{Cory}},
  \bibinfo{journal}{e-print: quant-ph/0303115}  (\bibinfo{year}{2003}).



\bibitem{foot2}
 For an arbitrary, quasi-static bath ({\em i.e.}, not
  necessary a spin-bath) with a density matrix that is diagonal in the
  eigenbasis of $\Az$, $\Phi_{FID} = e^{-i \delta t}
  \int_{-\infty}^{\infty} d\Lambda \rho(\Lambda) e^{-i \Lambda t}$,
  demonstrating that $\Phi_{FID}$ is exactly the inverse Fourier
  transform of the bath degree of freedom in this case.  

\bibitem{foot3}
By assuming the bath density matrix is diagonal in the $\Az$ eigenbasis, the result derived (Eqn.~\ref{e:finalRabiReal}) in fact is generally true for any bath that is non-singular ($\rho_{sym}(\omega \ge \Omega)$ not singular) and satisfies $u \ge 0$, not just a spin-bath.  However, the spin-bath provides a natural case for $[\hat H_B,\Az] \simeq 0$, as mentioned in the text.

\bibitem{foot4}
Well-separated singularities in $\rho_{sym}$ can be
  treated as additional stationary phase integral terms, and for each,
  corresponding oscillations at the resonance with different
  time-scales $u_j$ will emerge.

\bibitem[{\citenamefont{Gardiner}(1985)}]{gardiner85}
\bibinfo{author}{\bibfnamefont{C.~W.} \bibnamefont{Gardiner}},
  \emph{\bibinfo{title}{Handbook of stochastic methods}}
  (\bibinfo{publisher}{Berlin: Spinger}, \bibinfo{year}{1985}),
  \bibinfo{edition}{2nd} ed.

\bibitem[{\citenamefont{Mehring}(1976)}]{mehring76}
\bibinfo{author}{\bibfnamefont{M.}~\bibnamefont{Mehring}},
  \emph{\bibinfo{title}{High Resolution NMR Spectroscopy in Solids}}
  (\bibinfo{publisher}{Berlin: Springer-Verlag}, \bibinfo{year}{1976}).

\bibitem[{\citenamefont{Paget et~al.}(1977)\citenamefont{Paget, Lampel,
  Sapoval, and Safarov}}]{paget}
\bibinfo{author}{\bibfnamefont{D.}~\bibnamefont{Paget}},
  \bibinfo{author}{\bibfnamefont{G.}~\bibnamefont{Lampel}},
  \bibinfo{author}{\bibfnamefont{B.}~\bibnamefont{Sapoval}}, \bibnamefont{and}
  \bibinfo{author}{\bibfnamefont{V.}~\bibnamefont{Safarov}},
  \bibinfo{journal}{Phys. Rev. B} \textbf{\bibinfo{volume}{15}},
  \bibinfo{pages}{5780} (\bibinfo{year}{1977}).

\bibitem[{\citenamefont{Waugh et~al.}(1968)\citenamefont{Waugh, Huber, and
  Haeberlen}}]{whh}
\bibinfo{author}{\bibfnamefont{J.}~\bibnamefont{Waugh}},
  \bibinfo{author}{\bibfnamefont{L.}~\bibnamefont{Huber}}, \bibnamefont{and}
  \bibinfo{author}{\bibfnamefont{U.}~\bibnamefont{Haeberlen}},
  \bibinfo{journal}{Phys. Rev. Lett.} \textbf{\bibinfo{volume}{20}},
  \bibinfo{pages}{180} (\bibinfo{year}{1968}).

\bibitem[{\citenamefont{Kautz and Martinis}(1990)}]{kautz90}
\bibinfo{author}{\bibfnamefont{R.}~\bibnamefont{Kautz}} \bibnamefont{and}
  \bibinfo{author}{\bibfnamefont{J.}~\bibnamefont{Martinis}},
  \bibinfo{journal}{Phys. Rev. B} \textbf{\bibinfo{volume}{42}},
  \bibinfo{pages}{9903} (\bibinfo{year}{1990}).

\bibitem[{\citenamefont{Galperin and Gurevich}(1991)}]{galperin91}
\bibinfo{author}{\bibfnamefont{Y.~M.} \bibnamefont{Galperin}} \bibnamefont{and}
  \bibinfo{author}{\bibfnamefont{V.~L.} \bibnamefont{Gurevich}},
  \bibinfo{journal}{Phys. Rev. B} \textbf{\bibinfo{volume}{43}},
  \bibinfo{pages}{12900} (\bibinfo{year}{1991}).

\bibitem[{\citenamefont{Caldeira and Leggett}(1983)}]{caldeira83}
\bibinfo{author}{\bibfnamefont{A.~O.} \bibnamefont{Caldeira}} \bibnamefont{and}
  \bibinfo{author}{\bibfnamefont{A.~J.} \bibnamefont{Leggett}},
  \bibinfo{journal}{Ann. of Phys.} \textbf{\bibinfo{volume}{149}},
  \bibinfo{pages}{347} (\bibinfo{year}{1983}).

\bibitem[{\citenamefont{Chattah et~al.}(2003)\citenamefont{Chattah, lvarez,
  Levstein, Cucchietti, Pastawski, Raya, and Hirschinger}}]{chattah03}
\bibinfo{author}{\bibfnamefont{A.~K.} \bibnamefont{Chattah}},
  \bibinfo{author}{\bibfnamefont{G.~A.} \bibnamefont{lvarez}},
  \bibinfo{author}{\bibfnamefont{P.~R.} \bibnamefont{Levstein}},
  \bibinfo{author}{\bibfnamefont{F.~M.} \bibnamefont{Cucchietti}},
  \bibinfo{author}{\bibfnamefont{H.~M.} \bibnamefont{Pastawski}},
  \bibinfo{author}{\bibfnamefont{J.}~\bibnamefont{Raya}}, \bibnamefont{and}
  \bibinfo{author}{\bibfnamefont{J.}~\bibnamefont{Hirschinger}},
  \bibinfo{journal}{Journal of Chemical Physics}
  \textbf{\bibinfo{volume}{119}}, \bibinfo{pages}{7943} (\bibinfo{year}{2003}).

\bibitem[{\citenamefont{Danieli et~al.}(2005)\citenamefont{Danieli, Pastawski,
  and \'Alvarez}}]{danieli05}
\bibinfo{author}{\bibfnamefont{E.~P.} \bibnamefont{Danieli}},
  \bibinfo{author}{\bibfnamefont{H.~M.} \bibnamefont{Pastawski}},
  \bibnamefont{and} \bibinfo{author}{\bibfnamefont{G.~A.}
  \bibnamefont{\'Alvarez}}, \bibinfo{journal}{Chem. Phys. Lett.}
  \textbf{\bibinfo{volume}{402}}, \bibinfo{pages}{88} (\bibinfo{year}{2005}).

\bibitem[{\citenamefont{Facchi et~al.}(2005)\citenamefont{Facchi, Tasaki,
  Pascazio, Nakazato, Tokuse, and Lidar}}]{fachi05}
\bibinfo{author}{\bibfnamefont{P.}~\bibnamefont{Facchi}},
  \bibinfo{author}{\bibfnamefont{S.}~\bibnamefont{Tasaki}},
  \bibinfo{author}{\bibfnamefont{S.}~\bibnamefont{Pascazio}},
  \bibinfo{author}{\bibfnamefont{H.}~\bibnamefont{Nakazato}},
  \bibinfo{author}{\bibfnamefont{A.}~\bibnamefont{Tokuse}}, \bibnamefont{and}
  \bibinfo{author}{\bibfnamefont{D.}~\bibnamefont{Lidar}},
  \bibinfo{journal}{Phys. Rev. A} \textbf{\bibinfo{volume}{71}},
  \bibinfo{pages}{022302} (\bibinfo{year}{2005}).

\bibitem[{\citenamefont{Imamolgu et~al.}(2003)\citenamefont{Imamolgu, Knill,
  Tian, and Zoller}}]{imamoglu03}
\bibinfo{author}{\bibfnamefont{A.}~\bibnamefont{Imamolgu}},
  \bibinfo{author}{\bibfnamefont{E.}~\bibnamefont{Knill}},
  \bibinfo{author}{\bibfnamefont{L.}~\bibnamefont{Tian}}, \bibnamefont{and}
  \bibinfo{author}{\bibfnamefont{P.}~\bibnamefont{Zoller}},
  \bibinfo{journal}{Phys. Rev. Lett.} \textbf{\bibinfo{volume}{91}},
  \bibinfo{pages}{017402} (\bibinfo{year}{2003}).

\bibitem[{\citenamefont{Arecchi et~al.}(1972)\citenamefont{Arecchi, Courtens,
  Gilmore, and Thomas}}]{arecchi72}
\bibinfo{author}{\bibfnamefont{F.~T.} \bibnamefont{Arecchi}},
  \bibinfo{author}{\bibfnamefont{E.}~\bibnamefont{Courtens}},
  \bibinfo{author}{\bibfnamefont{R.}~\bibnamefont{Gilmore}}, \bibnamefont{and}
  \bibinfo{author}{\bibfnamefont{H.}~\bibnamefont{Thomas}},
  \bibinfo{journal}{Phys. Rev. A} \textbf{\bibinfo{volume}{6}},
  \bibinfo{pages}{2211} (\bibinfo{year}{1972}),
  \urlprefix\url{http://80-link.aps.org.ezp1.harvard.edu/abstract/PRA/v6/p2211%
}.

\end{thebibliography}

\end{document}